\documentclass[letterpaper,twocolumn,accepted=2026-08-17]{quantumarticle}
\pdfoutput=1

\usepackage{graphicx}
\usepackage{dcolumn}

\usepackage{multirow}%
\usepackage{array}
\usepackage{booktabs}
\usepackage{tabularx}
\usepackage{xcolor}%
\usepackage{booktabs}%
\usepackage{listings}%

\usepackage{bm}
\usepackage{braket}
\usepackage{enumerate}
\usepackage{pbox}
\usepackage{hyperref}

\usepackage{algorithm,algorithmic}
\usepackage{mathtools, bigstrut}
\usepackage{float}
\usepackage[numbers,sort&compress]
{natbib}

\definecolor{codered}{HTML}{81221B}
\definecolor{codegreen}{HTML}{327463}
\definecolor{codepurple}{HTML}{70309f}
\definecolor{backcolour}{HTML}{F2F2F2}

\lstdefinestyle{codeStyle}{
    backgroundcolor=\color{black!5}, 
    basicstyle=\ttfamily\fontsize{8.3}{9}\selectfont, 
    keywordstyle=\color{codegreen}\bfseries, 
    commentstyle=\color{gray}, 
    stringstyle=\color{codered}, 
    numberstyle=\footnotesize\color{gray}\ttfamily, 
    identifierstyle=\color{black}, 
    showstringspaces=false, 
    breaklines=true, 
    frame=top bottom, 
    rulecolor=\color{gray}, 
    framerule=0.65pt, 
    numbers=none, 
    numbersep=5pt, 
    tabsize=4, 
    captionpos=b, 
    columns=fullflexible, 
    keepspaces=true,      
    morekeywords={*,...} 
}
\hypersetup{
    colorlinks=true,      
    linkcolor=codered,    
    citecolor=codered,    
    filecolor=codered,    
    urlcolor=codered      
}

\newcommand{\code}[1]{\mbox{\lstinline[basicstyle=\ttfamily,language=Python]|#1|}}

\newcommand{\bmhead}[1]{\vspace{0.25cm}\noindent\textbf{\\#1.}}

\begin{document}

\title{A General Framework for Gradient-Based Optimization of Superconducting Quantum Circuits using Qubit Discovery as a Case Study}%

\author{Taha Rajabzadeh}
\email{tahar@stanford.edu}
\affiliation{Department of Electrical Engineering, Stanford University, Stanford, CA 94305 USA}
\author{Alex Boulton-McKeehan*}
\affiliation{E. L. Ginzton Laboratory and the Department of Applied Physics, Stanford University, Stanford, CA 94305 USA}
\author{Sam Bonkowsky*}
\affiliation{Department of Physics, Stanford University, Stanford, CA 94305 USA}
\author{David I. Schuster}
\affiliation{E. L. Ginzton Laboratory and the Department of Applied Physics, Stanford University, Stanford, CA 94305 USA}
\author{Amir H. Safavi-Naeini}
\affiliation{E. L. Ginzton Laboratory and the Department of Applied Physics, Stanford University, Stanford, CA 94305 USA}

\begin{abstract}
Engineering the Hamiltonian of a quantum system is fundamental to achieving novel goals or better performance. Automating Hamiltonian design through gradient-based optimization can dramatically accelerate this process. However, computing the gradients of eigenvalues and eigenvectors of the Hamiltonian---a large, sparse matrix---with respect to system properties poses a significant challenge, especially for arbitrary systems. Superconducting quantum circuits offer substantial flexibility in Hamiltonian design, making them an ideal platform for this task. In this work, we present a comprehensive framework for the gradient-based optimization of superconducting quantum circuits, leveraging the \code{SQcircuit} software package. By addressing the challenge of calculating the gradient of the eigensystem for large, sparse Hamiltonians and integrating automatic differentiation within \code{SQcircuit}, our framework enables efficient and precise computation of gradients for various circuit properties and custom metrics, streamlining the optimization process. We apply this framework to discover qubit designs with superior performance metrics. The optimized circuits show improvements in a heuristic measure of gate count, upper bounds on gate speed, decoherence time, and resilience to noise and fabrication errors compared to existing qubits. While this methodology is showcased through qubit optimization and discovery, it is versatile and can be extended to tackle other optimization challenges in superconducting quantum hardware design.
\end{abstract}

\maketitle
\makeatletter
\insert\footins{\footnotesize\sffamily\hsize\columnwidth\@parboxrestore\noindent
*A.B. and S.B. contributed equally to this work.\vspace{0.5\baselineskip}}
\makeatother

\tableofcontents

\section{Introduction}\label{sec:intro}
Superconducting quantum circuits are a leading hardware platform for realizing quantum computing in a laboratory environment, and are strong candidates for achieving fault-tolerant quantum computation. Consequently, there has been tremendous effort in this field to utilize superconducting circuits to create qubits with long coherence times \cite{soft_zero_pi, two_cooper_pair}, build local couplers to enable efficient qubit interaction \cite{tunable_coupler_fluxonium, tunable_crosstalk}, and develop hybrid systems for quantum sensing and memory applications \cite{nathan, padl}. Additionally, significant efforts are focused on studying quantum error correction \cite{qec_rev, google_surface_code, surface_code, qec_wallraff} and autonomous quantum error correction \cite{AutoQEC, aqec_exp} in these hardware platforms.

These applications, among many others, involve increasing complexities in circuit design and require a comprehensive approach for analyzing larger and more intricate circuits. Fundamental studies \cite{SQcircuit, kerman, Smith2016, Burkard2004, Yurke1984, symplectic_geometric, algebraic, faddeev_quantization, devoret} have led to the development of open-source software packages \cite{SQcircuit, qucat, scqubit, scqubit_2, circuitq} for analyzing the Hamiltonian of superconducting quantum circuits. In this work, we use and extend the \code{SQcircuit} Python package \cite{SQcircuit}, which is able to efficiently analyze arbitrary circuits by employing an advantageous coordinate transformation.

As is often the case with quantum systems, as circuits become larger, it becomes not only harder to simulate them numerically but also more challenging to build intuition. Although existing packages have made the task of analysis easier, the design process, even for a relatively simple qubit such as the transmon, requires consideration of many parameters and constraints. This indicates a need to automate the design process of superconducting circuits via optimization over superconducting quantum hardware.

There has been significant progress in the field, including remarkable work applying optimization and machine learning techniques to superconducting quantum hardware design. For example, \cite{4-local} applied an evolutionary algorithm to design a 4-local coupler for flux qubits, and \cite{joint_optimization} used gradient-based joint optimization to discover fluxonium qubits with better gate properties. However, applying these methodologies to new problems often requires starting from scratch, as the obstacles and constraints can change with each new problem. Therefore, there is a need for a universal approach that can be applied to arbitrary superconducting quantum circuits.

We construct an optimization framework for arbitrary superconducting quantum hardware. Based on mature approaches to gradient-based optimization employed in the machine learning community, we establish a similar method with a user-friendly interface in the \code{SQcircuit} package. To demonstrate its applications, we use it to automatically discover circuits which perform well as single qubits. This approach can be easily extended to other problems in superconducting quantum hardware, due to the ability to optimize for virtually any differentiable loss function constructed from element values and the eigenvalues and eigenvectors of the circuit Hamiltonian.

Our primary contributions from this work are as follows:
\begin{enumerate}[i.]
    \item Extended \code{SQcircuit} with automatic differentiation to compute the gradient of arbitrary metrics with respect to circuit element values.
    \item Developed a robust optimization pipeline for superconducting quantum circuit discovery using BFGS optimization, requiring only the definition of an appropriate loss function for various optimization problems.
    \item Implemented automatic assignment of Hilbert space truncation numbers for each circuit mode, along with diagonalization convergence tests, to ensure thorough and reliable exploration of the search space. This approach addresses significant bottlenecks commonly encountered in the search process.
    \item Introduced metrics to compare distinct qubits based on their laboratory implementability and total gate count. Leveraged our pipeline to optimize over this objective and thoroughly explore the space of single-loop circuit topologies for candidate qubits.
\end{enumerate}

The remainder of this manuscript is structured as follows: In Section~\ref{sec:review}, we review the superconducting quantum circuit analysis and coordinate transformations used by \code{SQcircuit}, establishing the theoretical foundation for subsequent sections. In Section~\ref{sec:gradient}, we formulate the general optimization problem for superconducting circuits, emphasizing the importance of gradient-based optimization. We then explain the theory behind the automatic differentiation implemented in \code{SQcircuit}. In Section~\ref{sec:qubit_discovery}, we focus on the qubit discovery problem as a demonstration of automatic differentiation and the optimization pipeline, which easily extends to other discovery problems. We also discuss the metrics we use to frame qubit discovery as a search problem and provide a detailed analysis of the optimization results and the discovered qubits. Finally, we present our conclusions and an outlook in Section~\ref{sec:conclusion}.
\section{Superconducting Quantum Circuit Analysis Review}\label{sec:review}
In this section, we present the analysis of an arbitrary superconducting quantum circuit constructed with lumped elements, as described in \cite{SQcircuit}. The Hamiltonian of a circuit with $n_N+1$ nodes ($n_N$ degrees of freedom) and $n_L$ independent inductive loops (number of constraints) is described by 
\begin{align}\label{eq:circuitH}
    \hat{{H}} = &\frac{1}{2} {\hat{\bm{Q}}}^T {\bm{C}}^{-1}{\hat{\bm{Q}}} + \frac{1}{2} {\hat{\bm\Phi}}^T {\bm{L}}^{*} {\hat{\bm\Phi}} +\sum_{k\in \mathcal{S}_L}\left(\frac{\Phi_0}{2\pi}\frac{\bm{b}_k^T\bm{\varphi}_{\text{ext}}}{l_k}\right){\bm{w}}^T_k{\hat{\bm{\Phi}}} \nonumber \\
    &-\sum_{k\in \mathcal{S}_J} E_{J_k} \cos \left(\frac{2\pi}{\Phi_0}{\bm{w}}^T_k{\hat{\bm{\Phi}}}+\bm{b}_k^T\bm{\varphi}_{\text{ext}}\right),
\end{align}
where:
\begin{itemize}
    \item $\hat{\bm{Q}} = [\hat{Q}_1, \cdots, \hat{Q}_{n_N}]$ is an $n_N$-dimensional vector containing charge operators, and $\hat{\bm{\Phi}} = [\hat{\Phi}_1, \cdots, \hat{\Phi}_{n_N}]$ is an $n_N$-dimensional vector containing flux operators. The operators satisfy the canonical commutation relation $[\hat{\bm{\Phi}}, \hat{\bm{Q}}] = i\hbar\bm{1}$.
    \item $\bm{C}$ and $\bm{L}^*$ are $n_N \times n_N$ capacitance and susceptance matrices, respectively. Diagonal elements are the sum of capacitances and susceptances connected to a node, and off-diagonal elements are the negative values of capacitance and susceptance between nodes.
    \item $E_{J_k}$ is the junction energy of the Josephson junction at branch $k$.
    \item $\mathcal{S}_L$ and $\mathcal{S}_J$ are the sets of branches which contain inductors and Josephson junctions, respectively.
    \item $\bm{w}_k$ is an $n_N$-dimensional vector; for branch $k$ between nodes $i$ and $j$, the $m$th element is $\delta_{im} - \delta_{mj}$ for element $m \in \{1,2,\cdots,n_N\}$.
    \item $\bm{b}_k$ is an $n_L$-dimensional vector where the $l$th element is $1$ if the $k$th branch completes loop $l$, and $0$ otherwise.
    \item $\bm{\varphi}_{\text{ext}}$ is an $n_L$-dimensional vector that is related to external fluxes via  $\bm{\varphi}_{\text{ext}} = 2\pi \bm{\Phi}_{\text{ext}} / \Phi_0$. Here, $\bm{\Phi}_{\text{ext}} = [\phi_{\text{ext},1}, \cdots, \phi_{\text{ext},n_L}]$ is a vector of the external fluxes applied to each independent inductive loop in the circuit.
\end{itemize}
Quantum analysis of superconducting circuits becomes computationally intensive as the circuit size increases. Efficient numerical simulation necessitates coordinate transformations that output a sparse Hamiltonian and lead to rapidly decaying off-diagonal matrix elements. To meet these criteria, we transform the Hamiltonian of Equation~\eqref{eq:circuitH} by performing a canonical transformation of charge and flux
operators:
\begin{align}\label{eq:transformations}
\hat{\tilde{\bm{Q}}} = \bm{R}^{-1}\hat{\bm{Q}},\
\hat{\tilde{\bm{\Phi}}} = \bm{S}^{-1}\hat{\bm{\Phi}}.
\end{align}
The matrices $\bm{R}$ and $\bm{S}$, which are $n_N \times n_N$ real invertible matrices, are used in a canonical transformation such that the transformed charge and flux operators, $\hat{\tilde{\phi}}_m$ and $\hat{\tilde{Q}}_n$, satisfy the commutation relation $[\hat{\tilde{\phi}}_m, \hat{\tilde{Q}}_n] = \delta_{mn}i\hbar$. This necessitates the constraint of $\bm{S}^T = \bm{R}^{-1}$. The transformed Hamiltonian then takes the form:
\begin{align*}
    \hat{\tilde{H}} = &\frac{1}{2} \hat{\tilde{\bm{Q}}}^T \tilde{\bm{C}}^{-1}\hat{\tilde{\bm{Q}}} + \frac{1}{2} \hat{\tilde{\bm\Phi}}^T \tilde{\bm{L}}^{*} \hat{\tilde{\bm\Phi}}+\sum_{k\in \mathcal{S}_L}\left(\frac{\Phi_0}{2\pi}\frac{\bm{b}_k^T\bm{\varphi}_{\text{ext}}}{l_k}\right)\tilde{\bm{w}}^T_k\hat{\tilde{\bm{\Phi}}}\\
    &-\sum_{k\in \mathcal{S}_J} E_{J_k} \cos \left(\frac{2\pi}{\Phi_0}\tilde{\bm{w}}^T_k\hat{\tilde{\bm{\Phi}}}+\bm{b}_k^T\bm{\varphi}_{\text{ext}}\right).
\end{align*}
The corresponding matrices and vector transformations are:
\begin{align*}
\tilde{\bm{C}}^{-1} = \bm{R}^T\bm{C}^{-1}\bm{R},\
\tilde{\bm{L}}^{} = \bm{S}^T\bm{L}^{}\bm{S},\
\tilde{\bm{w}}_k^T = \bm{w}_k^T\bm{S}.
\end{align*}
As detailed in Appendix~B of \cite{SQcircuit}, matrices $\bm{S}$ and $\bm{R}$ can be chosen to arrange $\tilde{\bm{L}}^{}$ and $\tilde{\bm{C}}$ into a block diagonal structure shown below:
\begin{align*}
    \tilde{\bm{C}}=\begin{bmatrix}
    \bm{C}^\text{ha}  &   \bm{0}\\
    \bm{0}   &   \bm{C}^\text{ch}
    \end{bmatrix},\
    \tilde{\bm{L}}^{*}=\begin{bmatrix}
    \bm{L}^{*^\text{ha}}   &   \bm{0}\\
    \bm{0}   &   \bm{0}
    \end{bmatrix},
\end{align*}
where $\bm{C}^{\text{ha}}$ and $\bm{L}^{^{\text{ha}}}$ are $n_H{\times}n_H$ invertible diagonal matrices termed the ``harmonic'' components, and $\bm{C}^{\text{ch}}$ is a $n_C{\times}n_C$ symmetric matrix (not necessarily diagonal), termed the ``charge'' component. This leads to a division of the transformed charge and flux operators into harmonic and charge contributions:
\begin{align*}
        \hat{\tilde{\bm{Q}}}^T=[\hat{Q}_1^{\text{ha}}, \dots, \hat{Q}_{n_H}^{\text{ha}}| \hat{Q}_1^{\text{ch}}, \dots, \hat{Q}^{ch}_{n_C}],\\
        \hat{\tilde{\bm{\Phi}}}^T=[\hat{\phi}_1^{\text{ha}}, \dots, \hat{\phi}_{n_H}^{\text{ha}}| \hat{\phi}_1^{\text{ch}}, \dots, \hat{\phi}_{n_C}^{\text{ch}}].
\end{align*}
The above transformation describes the system's dynamics in terms of (1) $n_H$ uncoupled harmonic oscillators with their dynamics defined by the diagonal elements of $\bm{C}^\text{ha}$ and $\bm{L}^{*^{\text{ha}}}$, (2) $n_C$ isolated superconducting islands with charging energy described by $\bm{C}^\text{ch}$, and (3) a flux distribution across various junctions determined by the vector $\tilde{\bm{w}}_k^T$. In representing the Hamiltonian, we then select the Fock basis for harmonic modes and the charge basis for charge modes and diagonalize the Hamiltonian in these respective bases. This entire process is automated by the \code{SQcircuit} software, which is described in more detail in \cite{SQcircuit}.
\section{Circuit Optimization and Gradient Calculation}\label{sec:gradient}
\begin{table*}[t]
\caption{Derivatives of the Hamiltonian in original and computational bases for various circuit parameters. These derivatives are computed by the custom computational node integrated into the \code{PyTorch} engine of \code{SQcircuit}.
}\label{tab:hamiltonian_gradient}%
\centering
\begin{tabular}{lll} \toprule
Hamiltonian Derivative & Original Basis & Computational Basis\\
\midrule
$\partial \hat{H}/\partial c_j$    & $-\frac{1}{2}\hat{\bm{Q}}^T \bm{C}^{-1} \frac{\partial\bm{C}}{\partial c_j}\bm{C}^{-1}\hat{\bm{Q}}$ & $ -\frac{1}{2}\hat{\tilde{\bm{Q}}}^T\bm{R}^T \bm{C}^{-1} \frac{\partial\bm{C}}{\partial c_j}\bm{C}^{-1}\bm{R}\hat{\tilde{\bm{Q}}}$ \\
\midrule
$\partial \hat{H}/\partial l_j$    & $\frac{1}{2}\hat{\bm{\Phi}}^T  \frac{\partial\bm{L}^*}{\partial l_j}\hat{\bm{\Phi}}$  & $\frac{1}{2}\hat{\tilde{\bm{\Phi}}}^T\bm{S}^T \frac{\partial\bm{L}^*}{\partial l_j}\bm{S}\hat{\tilde{\bm{\Phi}}}$ \\
\midrule
$\partial \hat{H}/\partial E_{J_j}$    & $- \cos \left(\frac{2\pi}{\Phi_0}{\bm{w}}^T_j{{\hat{\bm{\Phi}}}}+\bm{b}_j^T\bm{\varphi}_{\text{ext}}\right)$  & $-\cos \left(\frac{2\pi}{\Phi_0}\tilde{\bm{w}}^T_j\hat{\tilde{\bm{\Phi}}}+\bm{b}_j^T\bm{\varphi}_{\text{ext}}\right)$\\
\midrule
$\partial \hat{H}/\partial n_{g_j}$    & ${\bm{e}_j}^T {\bm{C}}^{-1}\hat{\bm{Q}}$ \footnotemark[1]   & ${\bm{e}_j}^T {\bm{C}}^{-1}\bm{R}\hat{\tilde{\bm{Q}}}$\\
\midrule
$\partial \hat{H}/\partial \varphi_{\text{ext}_j}$    & $\sum_{k\in \mathcal{S}_J} E_{J_k} b_{kj}\sin \left(\frac{2\pi}{\Phi_0}{\bm{w}}^T_k{{\hat{\bm{\Phi}}}}+\bm{b}_k^T\bm{\varphi}_{\text{ext}}\right)$ \footnotemark[2]& $\sum_{k\in \mathcal{S}_J} E_{J_k} b_{kj}\sin \left(\frac{2\pi}{\Phi_0}\tilde{\bm{w}}^T_j\hat{\tilde{\bm{\Phi}}}+\bm{b}_k^T\bm{\varphi}_{\text{ext}}\right)$ \\ \bottomrule
\end{tabular}
\footnotetext[1]{Here, $\bm{e}_j$ is an $n_N$-dimensional vector with its $j$-th element equal to $1$ and all other elements equal to $0$.}
\footnotetext[2]{We assume that all external fluxes are assigned to junctions, which implies that $\mathcal{S}_L$ is empty. This assumption holds true in the absence of a time-dependent Hamiltonian \cite{SQcircuit, time_dpendent_hamil_1, time_dpendent_hamil_2}.}
\end{table*}
\begin{figure}
    \centering
    \includegraphics[width=1\linewidth]{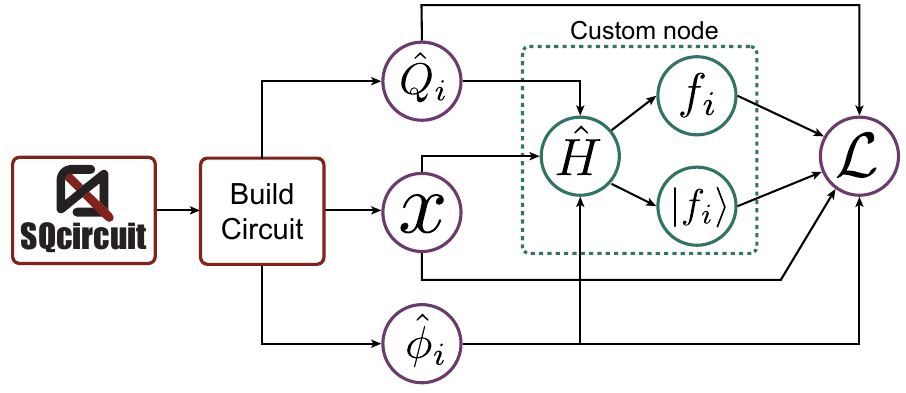}
    \caption{\textbf{Computational graph in \code{SQcircuit}.} This diagram illustrates the integration of a custom computational node into the \code{SQcircuit} framework, facilitating automatic differentiation of circuit properties. The process begins with construction of the circuit using \code{SQcircuit}. The charge operators $\hat{Q}_i$, flux operators $\hat{\Phi}_i$, and circuit elements $\bm{x}$ are implemented in \code{PyTorch} (represented above as purple nodes). Subsequently, eigenfrequencies $f_i$ and eigenvectors $\ket{f_i}$ are computed within a custom node (depicted in green). These results are then fed into the loss function $\mathcal{L}$, as defined in Equation~\eqref{eq:lossfunction}. This custom node performs the computation of gradients with respect to circuit elements $\bm{x}$, enabling the complete optimization process for superconducting circuit design within the \code{PyTorch} engine.}
    \label{fig:comp_graph}
\end{figure}
The optimization of superconducting quantum hardware systems can be framed in terms of minimizing a loss function, denoted as $\mathcal{L}$. This function quantifies the discrepancy between the properties of a simulated circuit and those of an ideal target circuit. A lower value of $\mathcal{L}$ indicates closer alignment with the desired circuit characteristics. Within this work, we consider loss functions of the form:
\begin{equation}\label{eq:lossfunction}
    \mathcal{L} = \mathcal{L}\left(\hat{Q}_i, \hat{\Phi}_i, f_i, \ket{f_i}, \bm{x}\right),
\end{equation}
where $\hat{Q}_i$ and $\hat{\Phi}_i$ represent the charge and flux operators at node $i$, respectively; $f_i$ and $\ket{f_i}$ denote the eigenfrequency and eigenvector of the $i^\text{th}$ mode, respectively; and $\bm{x}$ denotes a concatenated vector of all lumped circuit elements (capacitors, inductors, and junctions) and externally tunable parameters (gate charges $n_g$ and external fluxes $\varphi$).

Optimization of the objective then equates to finding the vector $\bm{x}$ for a given circuit topology that minimizes $\mathcal{L}$, constrained by the physical and fabrication limitations of the circuit element values. This can be formulated mathematically as:
\begin{equation}\label{eq:mainopt}
\begin{aligned}
\min_{\bm{x}} \quad & \mathcal{L}\left(\hat{Q}_i, \hat{\Phi}_i, f_i, \ket{f_i}, \bm{x}\right)\\
\textrm{s.t.} \quad & \bm{x}_\text{min} \le \bm{x} \le \bm{x}_\text{max},
\end{aligned}
\end{equation}
where $\bm{x}_\text{min}$ and $\bm{x}_\text{max}$ are the lower and upper bounds on the circuit elements, respectively. These bounds are dictated by the fabrication capabilities and physical constraints of the circuit components, which are discussed in greater detail in Section~\ref{sec:qubit_discovery}.

\bmhead{Importance of Gradient in Optimization}
Loss functions of the form \eqref{eq:mainopt} are not necessarily convex, so finding global optimum is difficult. However, it is feasible to find satisfactory local optima through the following two general approaches: 
\begin{enumerate}
    \item First- or second-order optimization methods; and
    \item Zero-order optimization methods.
\end{enumerate}
First-order and second-order optimization methods, such as gradient descent, Adam \cite{adam}, quasi-Newton \cite{BFGS,L-BFGS}, and Newton's method utilize first-order derivatives $\partial\mathcal{L}/\partial \bm{x}$ and second-order derivatives $\partial^2\mathcal{L}/\partial x_i x_j$ (the Hessian matrix) to solve the minimization problem. These algorithms are efficient in converging to local optima of the loss function but do not specifically target global optima. They are more suited for exploitation (efficient convergence to local optima) rather than exploration (sacrifice of computational resources to seek out better local optima).

In contrast, zero-order optimization methods (also known as black-box optimization) such as particle swarm optimization  (PSO) \cite{particle_swarm}, Bayesian optimization \cite{bayesian_optimization}, reinforcement learning \cite{reinforcement_1, reinforcement_2, reinforcement_3}, and evolutionary strategies \cite{evolutionary} attempt to solve the optimization problem without information about derivatives. These algorithms are advantageous for global optimum searches, albeit slower compared to first-order and second-order methods. They are more suitable for exploration, but require substantial computational resources to yield satisfactory results.

Generally, zero-order optimization is employed in scenarios where gradient calculations are either costly or impractical. However, supplementation of these algorithms with the local gradient can enhance their performance and balance exploration and exploitation more effectively. For instance, references \cite{particle_swarm_gradient} and \cite{bayesian_optimization_gradients} demonstrate improved performance in PSO and Bayesian optimization via incorporation of gradient information.

In conclusion, gradient information is highly valuable when it can be accurately computed or approximated, significantly enhancing the effectiveness of optimization algorithms in achieving more precise results or leading to simple and effective optimization when applied towards gradient-based techniques on their own. Within this work, we demonstrate how gradient-based optimization can be implemented straightforwardly and applied to achieve satisfactory results on its own.

\bmhead{Automatic Differentiation}
Our goal is to compute the gradient of an arbitrary loss function with respect to individual circuit parameters and elements. Analytically evaluating the gradient of the loss function is laborious and must be recalculated whenever the target problem changes, making manual computation impractical. However, the machine learning community has invested significant effort into automating the differentiation and gradient computation processes for diverse loss functions, originally designed for development of AI models. This need has led to the development of comprehensive automatic differentiation packages such as \code{PyTorch} \cite{pytorch}, \code{JAX} \cite{jax}, and \code{TensorFlow} \cite{tensorflow}. For instance, \code{PyTorch} calculates gradients using a `computational graph,' a network of nodes that represent variables and operations defining a function. The `forward pass' involves execution of these operations, applying a series of computational operations to input to produce output. Each operation yields intermediate results, forming a graph that traces the progression from input to output. During the `backward pass,' \code{PyTorch} traverses this graph in reverse, from outputs to inputs, applying the chain rule to efficiently propagate gradients backwards and facilitate input parameter optimization. The details of this gradient computation in \code{PyTorch} are illustrated in Appendix~\ref{app:PyTorchExample}.

In this work, we have integrated the \code{PyTorch} engine into the \code{SQcircuit} package, while maintaining backward compatibility with prior versions \cite{SQcircuit}. This integration allows \code{SQcircuit}, in conjunction with \code{PyTorch}, to compute derivatives of any arbitrary loss function of the form \eqref{eq:lossfunction} with respect to circuit parameters by applying the following chain rule:
\begin{equation}\label{eq:loss_chain_rule}
    \frac{d \mathcal{L}}{d \bm{x}} = \frac{\partial \mathcal{L}}{\partial \bm{x}} + \sum_i\frac{\partial f_i}{\partial \bm{x}}\frac{\partial \mathcal{L}}{\partial f_i} + \sum_i\frac{\partial \ket{f_i}}{\partial\bm{x}}\frac{\partial \mathcal{L}}{\partial \ket{f_i}}.
\end{equation}
Note that the charge and flux operators, represented as $\hat{Q}_i$ and $\hat{\Phi}_i$, do not depend on the circuit elements, hence $\partial \hat{Q}_i/\partial \bm{x} = \partial \hat{\Phi}_i/\partial \bm{x} = 0$. However, we cannot directly implement \eqref{eq:loss_chain_rule} using \code{PyTorch}'s internal functionalities since the gradients of eigenvalues and eigenvectors for sparse matrices are not currently supported. This feature is crucial for \code{SQcircuit}, which computes the Hamiltonian in a highly sparse basis to perform faster diagonalization.

To address this limitation, we introduce a custom  node into the computational graph (shown in Figure~\ref{fig:comp_graph}) that calculates the gradients of the eigenvalues and eigenvectors independently, and integrates these calculations with the \code{PyTorch} engine. This is achieved using perturbation theory, which approximates the derivatives of the eigenvalues and eigenvectors with respect to a parameter $x$ as follows:
\begin{align}
    \frac{\partial f_i}{\partial x} &= \langle f_i | \frac{\partial \hat{H}}{\partial x} | f_i \rangle, \label{eq:eig_val_grad}\\
    \frac{\partial | f_i \rangle}{\partial x} &= \sum_{m \neq i}^{n_E} \frac{\langle f_m | \frac{\partial \hat{H}}{\partial x} | f_i \rangle}{f_i - f_m} | f_m \rangle \label{eq:eig_vec_grad},
\end{align}
where $\hat{H}$ is the Hamiltonian defined in \eqref{eq:circuitH}, $n_E$ specifies the number of eigenvalues and eigenvectors calculated in the diagonalization process, and $x$ represents some circuit element or property such as $E_{J_j}$ (the junction at the $j$th branch), $c_j$ (the capacitor at the $j$th branch), $l_j$ (the inductor at the $j$th branch), $n_{g_j}$ (the gate charge number at the $j$th charge island), or $\varphi_{\text{ext}_j}$ (the external flux through the $j$th inductive loop).

It is important to note that the user cannot choose an arbitrarily large value for $n_E$, because the computational cost of the circuit diagonalization is proportional to $n_E$. However, $n_E$ must be sufficiently large to ensure adequate accuracy in the eigenvector gradient calculation. Calculating the gradient of the eigenvalues and eigenvectors thus requires determining $\partial \hat{H} /\partial x$ based on \eqref{eq:circuitH}. However, as \code{SQcircuit} represents the Hamiltonian and operators in the charge and harmonic computational bases, it is necessary to transform these operators from the original bases to the computational bases using the transformations defined in \eqref{eq:transformations}. Gradients of the Hamiltonian in both the original and computational bases are summarized in Table \ref{tab:hamiltonian_gradient}. 

We provide an example of using \code{SQcircuit} to calculate these gradients in Appendix~\ref{app:Tutorial}, showcasing the implementation of the computational graph in Figure~\ref{fig:comp_graph}.

In certain scenarios, it is necessary to compute the gradients of eigenfrequencies up to the second order. This requirement arises during the calculation of dephasing rates or $T_\varphi$ times, as they are themselves proportional to the gradient of the eigenfrequencies, as detailed in \cite{SQcircuit}. Consequently, these gradients are dependent on the second-order gradients of the eigenfrequencies. Due to the utilization of a custom calculation node within \texttt{PyTorch}, the second-order gradients of the eigenfrequencies are not computed automatically. We address this limitation by implementing a separate custom node that combines analytical expressions for second-order Hamiltonian derivatives with first-order automatic differentiation, as detailed in Appendix~\ref{app:T2_gradient}.

\section{Qubit Discovery}\label{sec:qubit_discovery}
\begin{figure}
    \centering
    \includegraphics[width=1\linewidth]{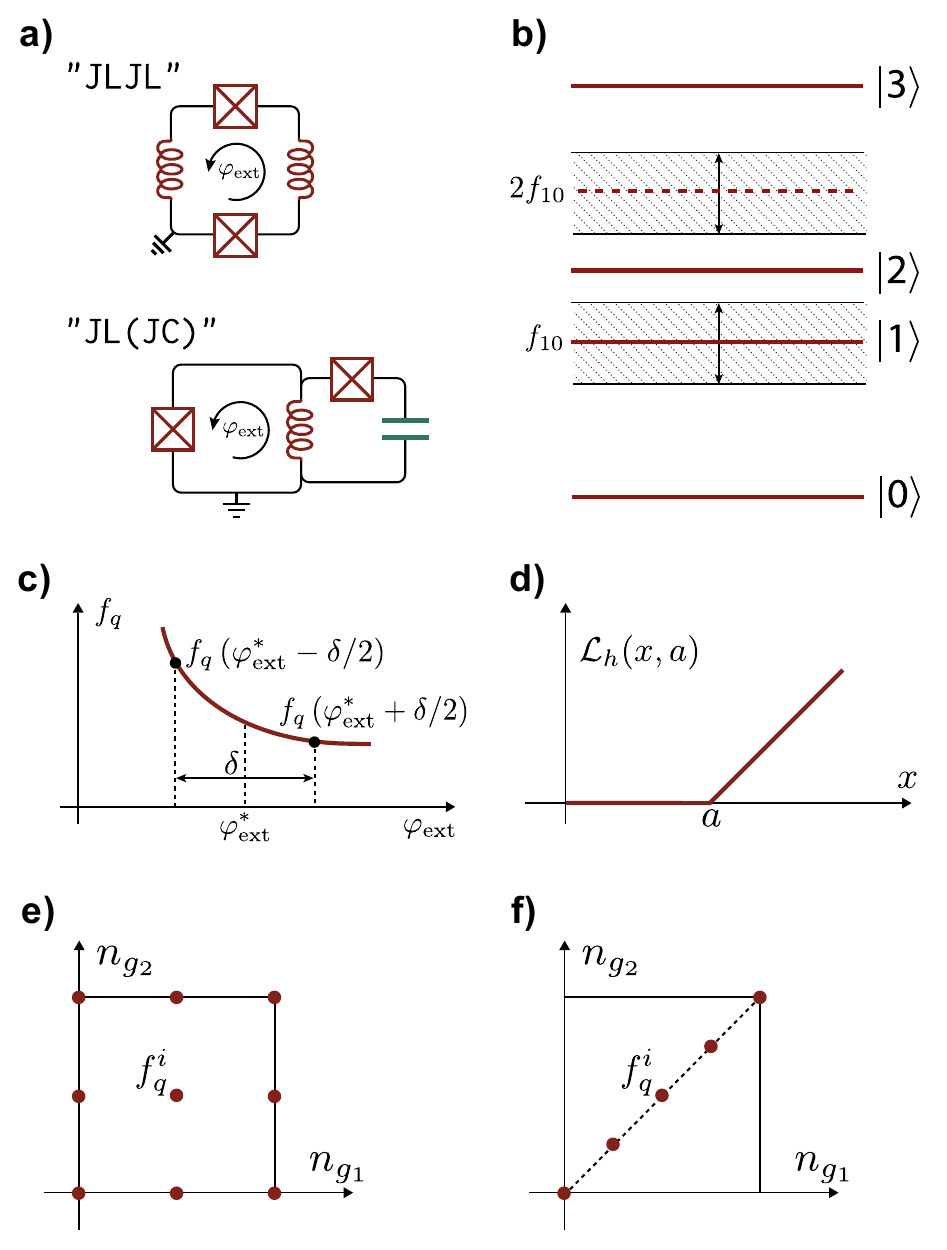}
    \caption{\textbf{Illustration of various qubit properties and metrics.} \textbf{a)} The circuit topology for the circuit codes \code{"JLJL"} and \code{"JL(JC)"} is presented. This circuit also includes an all-to-all capacitive connection; for the sake of simplicity, we omit these in the diagram. \textbf{b)} Energy-level diagram showing the transition frequencies between different qubit states. The dashed line represents twice the fundamental frequency $2f_{10}$, highlighting the transition caused by driving the qubit frequency $f_{10}$. The hatched box around these frequencies illustrates the sidebands, which are dependent on the gate speed. The edges of the box must not intersect with other states to prevent unwanted transitions from the qubit state to higher energy levels. Consequently, the maximum size of the hatched box defines the upper bound for the gate speed $\mathcal{G}$. \textbf{c)} Flux sensitivity $\mathcal{S}_\varphi$ as a function of external flux $\varphi_\text{ext}$, demonstrating the effect of flux variations on qubit frequency $f_q$. \textbf{d)} Hinge loss function $\mathcal{L}_h(x, a)$ used in the optimization problem to penalize constraint violations. \textbf{e)} Sampling grid for charge sensitivity $\mathcal{S}_n$ with two charge islands $n_{g_1}$ and $n_{g_2}$, showing sampled charge values. \textbf{f)} Uniform sampling along the diagonal of the hypercube for charge sensitivity calculation, demonstrating another method of charge sampling for circuits with two charge islands.}
    \label{fig:circuit_properties}
\end{figure}
\begin{figure*}
    \centering
    \includegraphics[width=1.0\linewidth]{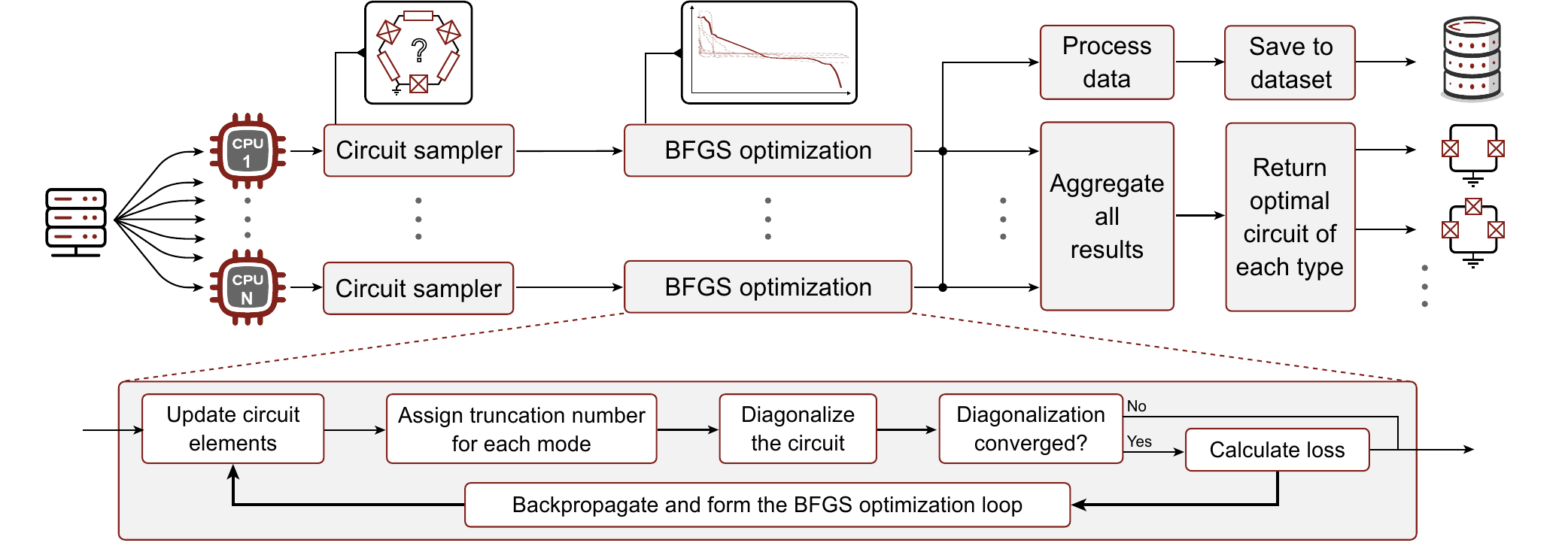}
    \caption{\textbf{Optimization pipeline.} The pipeline demonstrates the parallelized optimization process utilizing multiple CPU cores. Each core randomly initializes a superconducting quantum circuit through the circuit sampler module. The BFGS optimization algorithm is then applied, incorporating automatic truncation number assignment and convergence testing for robust optimization. The optimization loop iterates by calculating the loss and updating the circuit elements using gradient information and backpropagation. Aggregated results are processed to identify the optimal circuit configurations. Throughout the process, all circuits that pass the convergence test are stored and saved, creating a high-quality dataset useful for training AI and machine learning models.}
    \label{fig:optimization_pipeline}
\end{figure*}
The automatic differentiation capability of \code{SQcircuit}, discussed in Section~\ref{sec:gradient}, facilitates the study of many intriguing physical problems in superconducting circuit design that would be challenging to analyze without an appropriate tool. One of the main problems of interest is automating the process of designing useful and practical superconducting circuits. This issue is crucial because as superconducting circuits grow larger they become not only more difficult to simulate but also harder to understand intuitively. The number of circuit properties that designers must track also increases, further complicating the design process.

In this section, we address the problem of qubit discovery and automate the process by framing it as an optimization problem, which we solve using the autodifferentiation capability of \code{SQcircuit}. Though we have chosen to focus on qubit discovery in this section, our optimization pipeline and the techniques can also be applied to other superconducting quantum circuit discovery problems.

\subsection{Practical Qubit Design}
To evaluate and compare qubit designs, we need to establish criteria for their performance and feasibility. In the following, we describe the terminology and metrics used.
\bmhead{Search space ($\bm{x}_\text{min}$, $\bm{x}_\text{max}$)}
The element values for the circuit should fall within a range that can be fabricated. Therefore, we define the following conservative lower and upper bounds for capacitors, inductors, and junctions:
\begin{align}
    &c_\text{min} = 1~\text{femtofarad}, &c_\text{max} = 12~\text{picofarad}, \nonumber\\
    &l_\text{min} = 1~\text{femtohenry}, &l_\text{max} = 5~\text{microhenry}, \nonumber\\
    &E_{J_\text{min}} = 1~\text{gigahertz}, &E_{J_\text{max}} = 100~\text{gigahertz}. \nonumber
\end{align}
Accordingly, any search algorithm employed must ensure that the qubit components fall within these specified value ranges.
\bmhead{Circuit topology} 
The topology and arrangement of circuit elements within a graph fundamentally determines the behavior of superconducting circuits. This includes information about the circuit modes (either charge or harmonic), the control knobs for tuning the qubit, and the sources of noise that apply to the circuit. Consequently, this consideration is essential in our quest to identify the optimal qubit design. However, it is neither practical nor efficient to study all possible circuit topologies. Very large circuits are not only difficult to simulate, but also challenging to fabricate and test in a laboratory setting.

A complementary recent approach by Weissler \textit{et al.} systematically enumerates superconducting circuit topologies up to five nodes using graph-isomorphism methods, providing a useful catalogue of the broader circuit-design space~\cite{weissler2024enumeration}.

In this work, we then focus on circuits featuring a single  flux-tunable inductive loop and up to four nodes, with all-to-all capacitive coupling. Circuits without an inductive loop lack control knobs for tuning the qubit frequency, whereas circuits with more than one inductive loop become difficult to fabricate and tune. Limiting our study to circuits with up to four nodes ensures that the search space remains manageable with the computational resources at our disposal. An additional significant aspect of our selected topologies is that they encompass all qubits documented in the literature that have been experimentally realized. This allows us to use these well-established circuits as baselines for comparison with our optimized circuits.

Since our goal in this manuscript is to exhaustively study all possible circuit configurations, it becomes impractical to plot each circuit separately when referring to them. To address this, we have devised a coding scheme that represents individual circuit topologies as strings:
\begin{equation}
    \text{circuit code} = \texttt{"}\prod_{j=1}^{N} \mathcal{I}_j\left(\mathcal{E}_1^j\mathcal{E}_2^j\dots \mathcal{E}_{n_j}^j\right)\texttt{"},
\end{equation}
where $\mathcal{I}_j$ denotes either an inductor ($L$) or a Josephson junction ($J$) as an inductive element within a single inductive loop of the circuit. The parentheses next to the $\mathcal{I}_j$ element describes the non-inductive loop to which $\mathcal{I}_j$ belongs, and $\mathcal{E}_l^j$ represents the elements of that non-inductive loop, which can be either an inductor ($L$), a Josephson junction ($J$), or a capacitor ($C$). The parameter $n_j$ indicates the total number of elements in that non-inductive loop. It is important to note that at least one $\mathcal{E}_l^i$ must be a capacitor to prevent the formation of a new inductive loop. If $\mathcal{I}_j$ is not part of any non-inductive loop, the parentheses notation for that element is omitted. For example, the circuits with codes \code{"JLJL"} and \code{"JL(JC)"} are depicted in Figure~\ref{fig:circuit_properties}a.

As previously mentioned, our target topology search includes all known circuits from the literature. We summarize these circuits as follows: for \code{"JJ"}, we have the flux-tunable transmon from Google Sycamore \cite{sycamore}, for \code{"JL"}, we have the heavy fluxonium \cite{heavy_fluxonium} and Pop's fluxonium \cite{pop_fluxonium}, for \code{"JJJ"}, we have the flux qubit \cite{flux_qubit} and quantronium \cite{quantronium}, 
 and for \code{"JJL"}, we have the pokemon \cite{pokemon} and bifluxon \cite{bifluxon}. This collection serves as our baseline reservoir, against which we compare our discovery results.
\bmhead{Decoherence time ($T$)} 
The primary sources of decoherence in a qubit are depolarization and dephasing. Depolarization time $T_1$ is influenced by capacitive loss, inductive loss, and quasiparticle loss, while dephasing time $T_\varphi$ is affected by flux noise, charge noise, and critical current noise. The objective is to identify the qubit with the longest decoherence time, defined as:
\begin{equation}\label{eq:decoherence}
    T = \frac{1}{\frac{1}{2T_1} + \frac{1}{T_\varphi}}.
\end{equation}
This study considers all mentioned loss channels with an experimental basis defined in \cite{SQcircuit}. Such loss channels represent inherent environmental sources of noise and statistical fluctuations that affect qubits. In contrast, the flux and charge sensitivities help address crucial experimental considerations such as changes in qubit properties due to accumulation of electrons on charge islands as well as control of applied external flux at the critical flux point.

\bmhead{Single-qubit gate speed ($\mathcal{G}$)}
Another crucial metric for a qubit is the speed at which a single-qubit gate operation can be performed. Although this value generally depends on a time-dependent control signal, this work neglects the drive signal and instead calculates an upper bound for the maximum single-qubit gate speed under ideal conditions. This upper bound is determined by the largest sideband frequency around an integer multiple of the qubit frequency that does not excite the qubit out of its computational states.

We depict this consideration in Figure~\ref{fig:circuit_properties}b, and provide further discussion in Appendix~\ref{app:gate_speed}. The mathematical expression for the upper bound is given by:
\begin{equation}\label{eq:gate_speed}
    \mathcal{G} = \min_{i \geq 2} \left( f_i - f_1,\left| f_i - 2f_1 \right|\right),
\end{equation}
where $f_i$ represents the frequency of the $i$-th mode. Specifically, for $i=2$, the second term reduces to the standard definition of anharmonicity $\mathcal{A}$ within the literature. This demonstrates the generality of our definition, as it is shown that for fluxonium and transmon, the fastest single-qubit gate with optimized drive signals equals the anharmonicity.
\bmhead{Number of single-qubit gates ($\mathcal{N}$)}
The number of single-qubit gates that can be performed serves as a principal metric for comparing different qubits and benchmarking them against those reported in the literature. 

This number can be approximated by determining how many sequential gates can be executed as quickly as possible before the qubit loses its information to the environment, which we express by
\begin{equation}\label{eq:gate_number}
\mathcal{N} = T\mathcal{G},
\end{equation}
where $T$ represents the decoherence time, and $\mathcal{G}$ is the gate speed upper bound defined previously. This metric aims to strike a balance between maximizing the decoherence time and minimizing the gate operational time, making it a suitable unitless metric for optimization purposes.
\bmhead{Flux sensitivity ($\mathcal{S}_\varphi$)}
External fluxes coupling to the inductive loops of superconducting circuits serve as an important and practical degree of freedom in the lab to tune the qubit frequency to its desired working point. However, if the external flux working point $\varphi_\text{ext}^*$ is too sensitive to external flux variations, noise fluctuations in the lab can compromise qubit operation. We must then ensure that flux fluctuations due to noise remain within a tolerable regime.

Assuming $\delta$ as the maximum range of fluctuation, we define flux sensitivity as (see Figure~\ref{fig:circuit_properties}c):
\begin{equation}\label{eq:flux_sensitivity}
    \mathcal{S}_\varphi = \frac{\left| f_q\left(\varphi_\text{ext}^* + \frac{\delta}{2}\right) - f_q\left(\varphi_\text{ext}^* - \frac{\delta}{2}\right) \right|}{f_q\left(\varphi_\text{ext}^* \right)},
\end{equation}
where $f_q(.)$ is the qubit frequency as a function of external flux.
\bmhead{Charge sensitivity ($\mathcal{S}_n$)}
The charge on charge islands can be controlled using DC voltages, offering another degree of freedom for qubit control. However, charge dispersion in qubits is not practical because charge islands accumulate charges over time. To maintain a stable qubit, we must eliminate charge dispersion.

We define charge sensitivity by sampling charge values across all charge degrees of freedom (only charge modes), then calculating the ratio of the maximum qubit frequency range to the average of the minimum and maximum qubit frequency for these samples:
\begin{equation}\label{eq:charge_sensitivity}
    \mathcal{S}_n = \frac{\max_i f_q\left(\bm{n}_g^i\right) - \min_i f_q\left(\bm{n}_g^i\right)}{\frac{1}{2}\left(\max_i f_q\left(\bm{n}_g^i\right) + \min_i f_q\left(\bm{n}_g^i\right)\right)},
\end{equation}
where $f_q(.)$ is the qubit frequency as a function of charge numbers on the islands, and $\bm{n}_g^i = \left(n_{g_1}^i, n_{g_2}^i, \dots \right)$ is a vector of charge numbers for the $j$th island $n_{g_j}^i$ in the $i$th sample. Sampling can be performed over a grid (see Figure~\ref{fig:circuit_properties}e for a circuit with two charge islands) or uniformly across the diagonal of the hypercube formed by charge island degrees of freedom (see Figure~\ref{fig:circuit_properties}f for a circuit with two charge islands), the latter being the method used in this work.
\bmhead{Element sensitivity ($\mathcal{S}_e$)}
Another fundamental factor to consider is the qubit sensitivity to fabrication errors. A qubit might exhibit good metrics, but if it is highly sensitive to fabrication errors, achieving a practical yield of this qubit may be challenging. Consequently, this qubit may not provide the expected number of gates.

To account for this effect, we first sample our element values from a normal distribution. For the $i$th element with value $x_i$, we sample from a normal distribution with a mean of $x_i$ and a standard deviation of $e_i \cdot x_i$, where $e_i$ represents the fabrication error of the $i$th element. By sampling our circuits $n_S$ times, we generate $n_S$ circuits, each with a different number of gates. We define the element sensitivity as the standard deviation of the calculated gate numbers divided by their average, expressed as:
\begin{equation}\label{eq:element_sensitivity}
    \mathcal{S}_e = \frac{\text{std}\{\mathcal{N}\}}{\text{mean}\{\mathcal{N}\}}.
\end{equation}
\subsection{Search Problem as an Optimization} 
\begin{table*}[t]
\centering
\caption{\textbf{Qubit Discovery Results.} The table below presents the discovered qubits along with their corresponding circuit codes, metrics, and performance characteristics. The metrics include the total loss function $\mathcal{L}$ \eqref{eq:qubit_search_2}, the number of gates $\mathcal{N}$ \eqref{eq:gate_number}, the gate speed $\mathcal{G}$ \eqref{eq:gate_speed}, the decoherence time $T$ \eqref{eq:decoherence}, and the sensitivities $\mathcal{S}_\varphi$ \eqref{eq:flux_sensitivity}, $\mathcal{S}_n$ \eqref{eq:charge_sensitivity}, and $\mathcal{S}_e$ \eqref{eq:element_sensitivity}. Additionally, the qubit frequency $f_q$ and the external flux working point $\varphi^*_\text{ext}/2\pi$ are listed. Optimized qubits are compared with existing qubits from the literature to highlight the improvements achieved through the optimization pipeline.
}
\label{tab:search_results}
\small
\setlength{\tabcolsep}{4pt}
\begin{tabular}{ l >{\ttfamily}l >{\ttfamily}l >{\ttfamily}l >{\ttfamily}l >{\ttfamily}l >{\ttfamily}l >{\ttfamily}l >{\ttfamily}l >{\ttfamily}l} \toprule
\textbf{Circuit Code} & $\mathcal{L}$ & $\mathcal{N}$ & $\mathcal{G}$ [GHz] & $T$ [s] & $\mathcal{S}_\varphi$ & $\mathcal{S}_n$ & $\mathcal{S}_e$ & $f_q$ [GHz] & $\varphi^*_\text{ext}/2\pi$ \\
\midrule
\code{"JJ 1"} & 1.84E-05 & 5.43E+04 & 0.441 & 1.23E-04 & 1.08E-03 & 2.00E-02 & 5.78E-03 & 1.626 & 0.0200 \\ 
\code{"JJ 2"} & 2.09E-05 & 4.77E+04 & 1.167 & 4.09E-05 & 2.18E-04 & 2.00E-02 & 4.15E-02 & 4.305 & 0.4900 \\ 
Sycamore \cite{sycamore}& 2.32E-04 & 4.32E+03 & 0.191 & 2.25E-05 & 7.61E-04 & 8.61E-14 & 6.87E-03 & 7.013 & 0.0100 \\ 
\midrule
\code{"JL 1"} & 2.89E-07 & 3.46E+06 & 3.605 & 9.60E-04 & 1.35E-02 & 0.00E+00 & 1.99E-02 & 0.192 & 0.4990 \\
\code{"JL 2"} & 6.97E-07 & 1.43E+06 & 2.493 & 5.76E-04 & 1.99E-03 & 0.00E+00 & 2.28E-02 & 0.535 & 0.4900 \\
Heavy fluxonium \cite{heavy_fluxonium}& 2.56E+00 & 6.38E+05 & 2.937 & 2.17E-04 & 2.66E+00 & 0.00E+00 & 4.35E-02 & 0.014 & 0.4999 \\
Pop fluxonium \cite{pop_fluxonium}& 1.44E-06 & 6.94E+05 & 10.32 & 6.73E-05 & 3.35E-02 & 0.00E+00 & 2.43E-02 & 0.639 & 0.5000 \\ 
\midrule
\code{"JJJ"}& 1.10E-05 & 9.13E+04 & 1.231 & 7.42E-05 & 1.06E-03 & 8.62E-13 & 5.99E-03 & 2.596 & 0.0200 \\
Flux qubit \cite{flux_qubit}& 1.15E-04 & 8.68E+03 & 0.406 & 2.14E-05 & 1.70E-02 & 2.89E-09 & 3.33E-02 & 4.530 & 0.0500 \\ 
Quantronium~\cite{quantronium} & 1.90E-01 & 3.86E+04 & 7.520 & 5.14E-06 & 1.24E-03 & 2.10E-01 & 9.62E-03 & 16.204 & 0.0100 \\ 
\code{"JJ(JC)"} & 8.74E-06 & 1.14E+05 & 0.202 & 5.67E-04 & 7.37E-10 & 2.00E-02 & 1.04E-02 & 0.745 & 0.4200 \\ 
\midrule
\code{"JJL"} & 1.50E-06 & 6.66E+05 & 1.401 & 4.75E-04 & 1.09E-02 & 2.32E-10 & 4.33E-02 & 0.093 & 0.4990 \\ 
Pokemon \cite{pokemon}& 9.79E+01 & 7.67E+03 & 3.768 & 2.04E-06 & 9.80E+01 & 3.71E-08 & 9.82E-03 & 0.002 & 0.4999 \\ 
\code{"JL(JC)"} & 1.10E-06 & 9.06E+05 & 0.903 & 1.00E-03 & 1.24E-03 & 3.41E-11 & 1.16E-02 & 0.210 & 0.4990 \\ 
\midrule
\code{"JLL"} & 4.31E-07 & 2.32E+06 & 7.643 & 3.04E-04 & 9.68E-04 & 1.00E-14 & 2.11E-02 & 0.802 & 0.4990 \\ 
\midrule
\code{"JJJJ"} & 6.62E-06 & 1.51E+05 & 0.899 & 1.68E-04 & 2.78E-05 & 3.32E-04 & 1.36E-02 & 1.891 & 0.4990 \\ 
\midrule
\code{"JJJL"} & 1.44E-06 & 6.95E+05 & 0.853 & 8.15E-04 & 2.10E-03 & 2.74E-04 & 4.69E-03 & 0.163 & 0.4990 \\ 
\midrule
\code{"JLJL"} & 1.28E-06 & 7.82E+05 & 1.646 & 4.75E-04 & 2.03E-03 & 5.41E-06 & 5.74E-03 & 0.269 & 0.4990 \\ \bottomrule
\end{tabular}
\end{table*}
\begin{figure}
    \centering
    \includegraphics[width=1\linewidth]{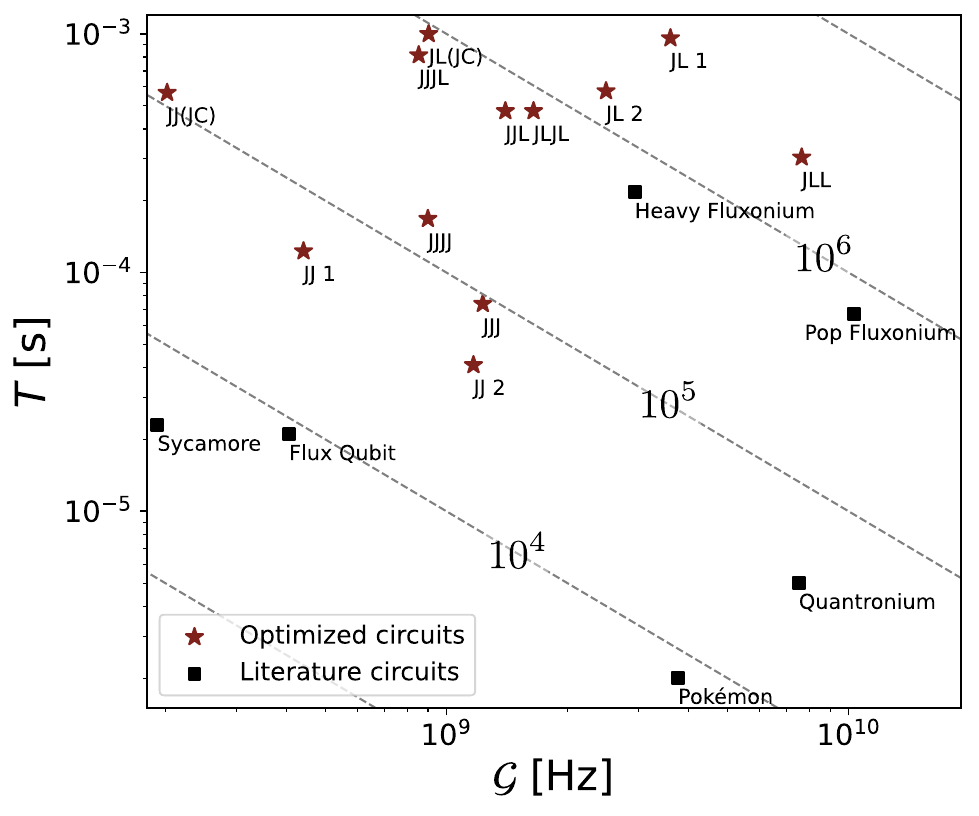}
    \caption{\textbf{Gate number of discovered qubits.} The qubits from Table~\ref{tab:search_results} plotted based on decoherence time $T$ and upper bound gate speed $\mathcal{G}$. The x-axis represents $\mathcal{G}$ and the y-axis represents $T$, both presented on a logarithmic scale. The dashed lines indicate constant contours of the gate number, following the expression $\mathcal{G}T=\text{const}$. Qubits towards the top right corner of the plot have a larger number of gates. Optimized qubits are marked with red stars, while qubits from the literature are marked with black squares.}
    \label{fig:search_results}
\end{figure}
\begin{figure*}
    \centering
    \includegraphics[width=1\linewidth]{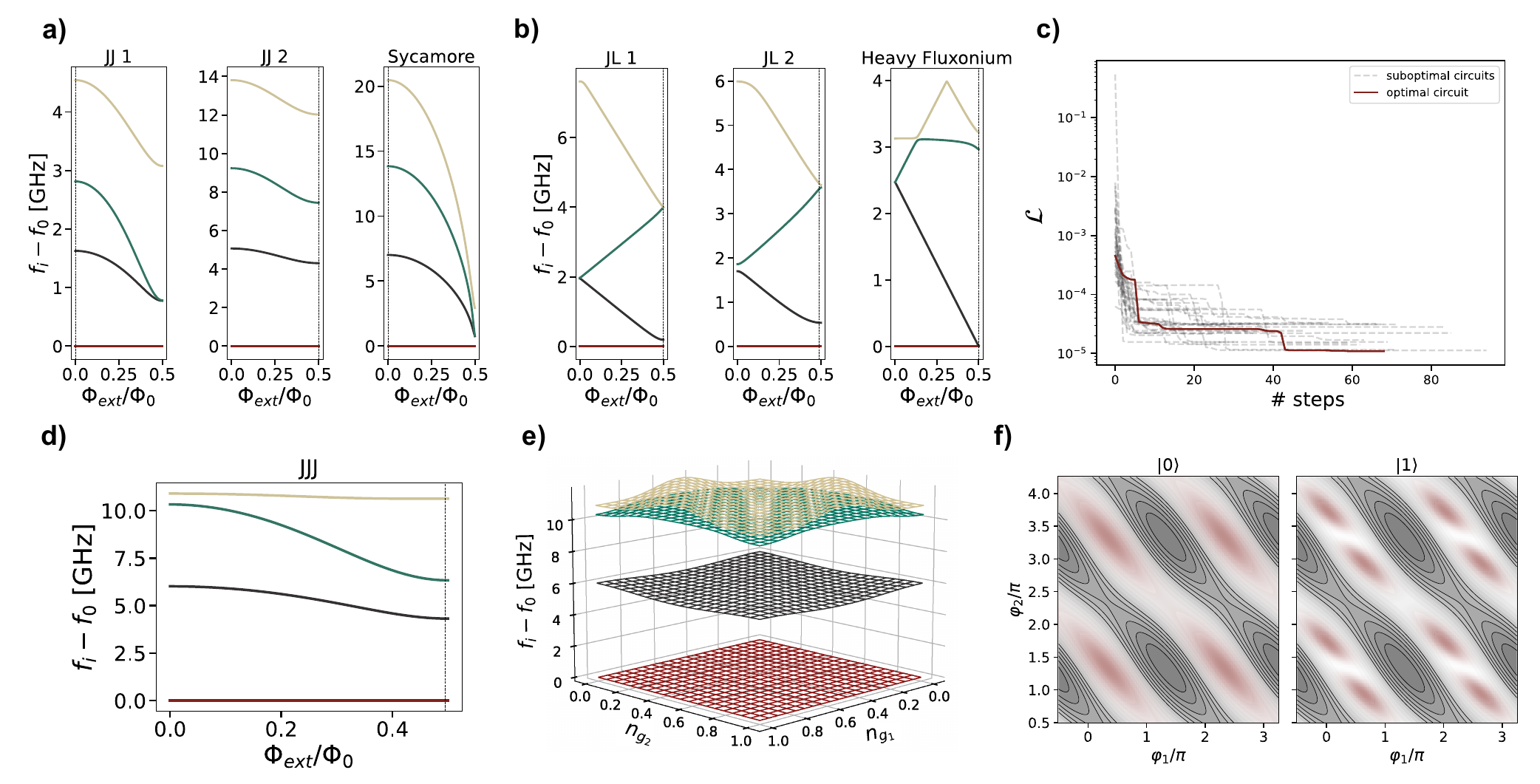}
    \caption{\textbf{Optimal circuit properties.} \textbf{a, b)} External flux spectrum of the discovered \code{"JJ"} and \code{"JL"} qubits compared to the Sycamore transmon and heavy fluxonium qubits. The external flux working points of each qubit are specified by a vertical dashed line. The element values for each qubit are specified in the Table~\ref{tab:JJ_values} and Table ~\ref{tab:JL_values}.
    \textbf{c)} BFGS optimization of one hundred instances of the \code{"JJJ"} qubit topology. Dashed black lines indicate suboptimal circuits, and the solid red line indicates the optimal circuit.
    \textbf{d)} Flux spectrum of the optimal \code{"JJJ"} circuit. The external flux working point is specified by a vertical dashed line.
    \textbf{e)} Charge spectrum for the excited states of the optimal \code{"JJJ"} qubit, showing insensitivity to charge for the first excited state, represented by the black surface.
    \textbf{f)} Eigenstates for the ground state and first excited state of the \code{"JJJ"} qubit in the Hamiltonian potential landscape, indicating two-dimensional transmon-like eigenvectors localized in the circuit potential minimum. The gray area with contours specifies the potential of the qubit, and the white area specifies the minimum of the potential. The red color shows the localized states.
}
    \label{fig:optimal_circuits}
\end{figure*}

\begin{figure*}
    \centering
    \includegraphics[width=1\linewidth]{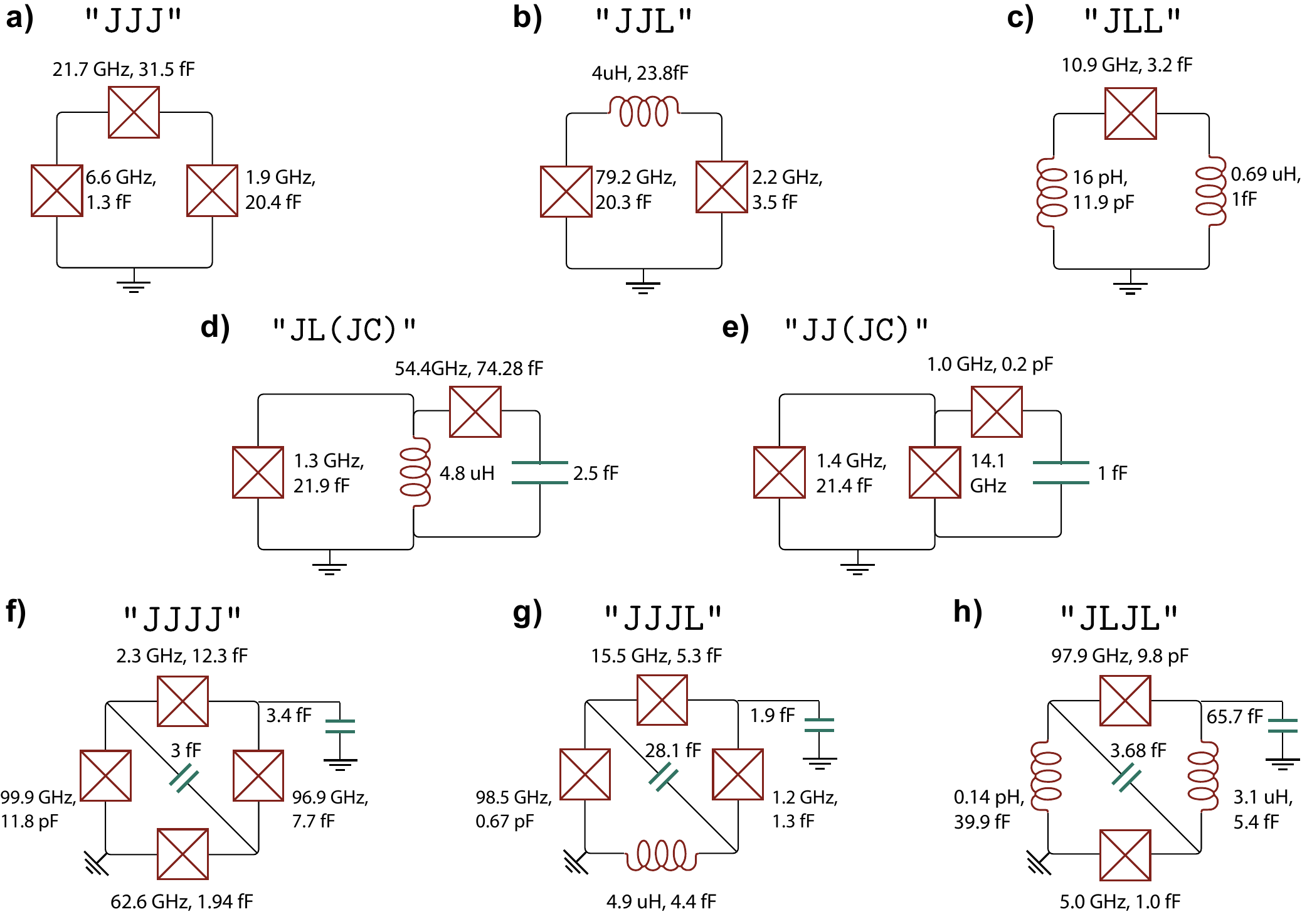}
    \caption{\textbf{Optimal circuit element values and their topologies.}
    \textbf{a)} \code{"JJJ"} qubit: Three Josephson junctions in an inductive loop with associated capacitors.
    \textbf{b)} \code{"JJL"} qubit: Two Josephson junctions and one inductor forming an inductive loop with associated capacitors.
    \textbf{c)} \code{"JLL"} qubit: Composed of two inductors and one junction, forming an inductive loop. Despite suggesting a large number of gates, it effectively reduces to fluxonium over the course of the optimization process.
    \textbf{d)} \code{"JL(JC)"} qubit: A fluxonium capacitively coupled to a Cooper pair box. The optimization process makes the fluxonium off-resonant with the Cooper pair box, causing it to dominate the lower levels of the system.
    \textbf{e)} \code{"JJ(JC)"} qubit: A flux-tunable transmon capacitively coupled to a non-flux-tunable Cooper pair box.
    \textbf{f--h)} Various other qubits such as \code{"JJJJ"}, \code{"JJJL"}, and \code{"JLJL"}, which present interesting qubit designs. Note while these circuits have all-to-all capacitive coupling, we only explicitly indicate the capacitors for branches without an inductive element.}
    \label{fig:circuits}
\end{figure*}

Now that we have defined all necessary metrics for evaluating and benchmarking a qubit, we can formulate our search for the best qubit as a maximization problem. This problem aims to maximize the number of gates while imposing constraints to ensure the qubits are implementable.

The optimization problem can be written as:
\begin{equation}\label{eq:qubit_search}
\begin{aligned}
\max_{\bm{x}} \quad & \mathcal{N} \\
\textrm{s.t.} \quad & \bm{x}_\text{min} \leq \bm{x} \leq \bm{x}_\text{max}, \\
              \quad & \mathcal{S}_i \leq \mathcal{S}_i^*, \quad i \in \{\varphi, n, e\}, \\
              \quad & f_q \leq f^*_q, \\
\end{aligned}
\end{equation}
where $\mathcal{S}_i^*$ is the maximum tolerance for the sensitivity of type $i$, which can be flux, charge, or element sensitivity, and $f^*_q$ is the maximum possible frequency imposed by the material from which the qubit is built. We propose a dual optimization problem whose solution is equivalent to that of the optimization problem in \eqref{eq:qubit_search}.

To this end, we define the optimization problem as:
\begin{equation}\label{eq:qubit_search_2}
\begin{aligned}
\min_{\bm{x}} \quad & \mathcal{L}_\text{obj} + \mathcal{L}_\text{const} \\
\textrm{s.t.} \quad & \bm{x}_\text{min} \leq \bm{x} \leq \bm{x}_\text{max},
\end{aligned}
\end{equation}
where
\begin{align*}
    &\mathcal{L}_\text{obj} = \frac{1}{\mathcal{N}}, \\ 
    &\mathcal{L}_\text{const} = \sum_{i \in \{\varphi, n, e\}} \beta_i \mathcal{L}_h(\mathcal{S}_i, \mathcal{S}_i^*) + \beta_f\mathcal{L}_h(\frac{f_q}{f_q^*}, 1),
\end{align*}
with $\beta_i$ being the loss function weights, and $\mathcal{L}_h(.,.)$ the hinge loss function depicted in Figure~\ref{fig:circuit_properties}d defined as:
\begin{equation*}
\mathcal{L}_h(x, a) = 
    \begin{cases}
      x - a & \text{if } x > a, \\
      0 & \text{if } x \leq a.
   \end{cases}
\end{equation*}
We use hinge loss to penalize the unsatisfiability of constraints in the optimization problem of Equation~\eqref{eq:qubit_search}, returning a zero gradient when each constraint is satisfied. Wrapping the sensitivity loss terms in a hinge loss ensures that once the flux and charge sensitivities are below some threshold, the model can divert its attention to focus on scalar loss terms such as the decoherence and gate speed.

In this work, we set all loss weights to $\beta_\varphi=\beta_n=\beta_e=\beta_f=1$. This choice is motivated by the fact that the constraint terms are normalized hinge losses, so unit weights assign comparable penalties to order-one violations of each constraint. Since the hinge penalties vanish once the constraints are satisfied, the final feasible optima are determined primarily by maximizing $\mathcal{N}=T\mathcal{G}$ rather than by fine tuning the $\beta_i$ values. In practice, we find that the results are not strongly sensitive to the precise choice of $\beta_i$ within roughly an order of magnitude around unity; outside this range, too-small values can allow constraint violations, while much larger values can make the optimization more conservative.

We denote $\bm{x}^\star$ as the global optimum of \eqref{eq:qubit_search} which satisfies each constraint and maximizes the gate number $\mathcal{N}^\star$. If the coefficients $\beta_i$ are sufficiently large, it is clear that $\bm{x}^\star$ is also the global optimum of \eqref{eq:qubit_search_2}. This is because $\mathcal{L}_\text{const} = 0$ (as all hinge losses are zero due to constraint satisfaction), and $\mathcal{N}^\star$ also minimizes $\mathcal{L}_\text{obj}$ by definition. Since the optimization problem in \eqref{eq:qubit_search_2} satisfies the form of \eqref{eq:mainopt}, we can use \code{SQcircuit} functionalities to perform gradient-based optimization and solve this problem.

\subsection{Optimization Pipeline} To solve the optimization problem in Equation~\eqref{eq:qubit_search_2}, we employ the BFGS algorithm \cite{BFGS} with random restarts. The BFGS algorithm is a quasi-Newton method that approximates the Hessian matrix of second-order partial derivatives of the loss function using updates derived from gradient evaluations. This approach avoids the need for a fixed step size, which is crucial for a fast and efficient optimization. The random restart strategy involves running multiple instances of BFGS with different circuit initializations and comparing their results to identify better local optima (or potentially the global optimum). By leveraging multiple CPU cores to execute these BFGS optimizations in parallel, we can significantly accelerate the search for the global optimum. The entire optimization pipeline is illustrated in Figure~\ref{fig:optimization_pipeline}, with details elaborated upon below.

The process begins by assigning the optimization tasks to $n_\text{CPU}$ CPU cores. Each core randomly initializes a superconducting quantum circuit via the circuit sampler module, within the range of elements and circuit topologies discussed earlier. 

We then assign a truncation number to each mode of the circuit. This altogether defines the total size of the Hilbert space $K$ and represents the main constraint upon available computational power and memory. Typically, equal truncation numbers are assigned to each mode due to the uncertainty of which mode requires more truncation. However, in this manuscript we employ our own method, detailed in Appendix~\ref{app:automatic_truncation}, which is more efficient than equal truncation and leads to more efficient circuit diagonalization.

In the subsequent step, we diagonalize the circuit and check for convergence of the eigenvalues and eigenvectors using the methodology described in Appendix~\ref{app:convergence_test}. If the circuit does not converge, we terminate the process and archive the circuit for further investigation. If the circuit does converge, we calculate the loss function and use automatic differentiation within \code{SQcircuit} to compute the gradient and run the optimization loop for BFGS. After all runs have completed, we aggregate the results and select the best circuit for that topology, as each topology involves distinct physics worth studying.

Our automatic truncation number assignment, combined with our convergence test, allows us to robustly explore the entire target search space and reliably diagonalize explored circuits with adequate accuracy. As a beneficial side effect, this process generates a collection of well-characterized circuits. While not the primary focus of our work, this data could potentially be used in future research, such as for training machine learning models to analyze superconducting quantum circuits.

As a final point, it is important to note that we do not directly optimize over unitful circuit element values $\bm{x}$. The typical units in which we choose to express the circuit element properties vary significantly over orders of magnitude. For example, capacitors can easily range from picofarads to femtofarads. This variation can make the loss landscape ill-conditioned for optimization. To address this issue, we apply a reparameterization technique to our problem, optimizing over $\bm{\alpha}_x$ rather than $\bm{x}$, with the following relationship:
\begin{equation*}
\begin{split}
    \log x = {}& \frac{\log x_\text{max} + \log x_\text{min}}{2} \\
    &+ \left( \frac{\log x_\text{max} - \log x_\text{min}}{2}\right)\cos \alpha_x.
\end{split}
\end{equation*}
This approach ensures that the values of $x$ remain within the specified range while optimizing over the unitless values $\bm{\alpha}_x$, which lie between $0$ and $2\pi$. The logarithmic scale aids in exploring the range of element values across orders of magnitude and prevents the optimizer from neglecting smaller element values.

\subsection{Discovered Qubits}
\begin{table}[t]
\centering
\centering
\setlength{\tabcolsep}{4pt}
\caption{\textbf{\code{"JJ"} element values.} Optimized circuit element values for \code{"JJ"} qubits compared to the Sycamore qubit. The table lists the Josephson junction energies ($E_{J_1}$ and $E_{J_2}$) and capacitance ($C$) for each circuit.}
\label{tab:JJ_values}
\begin{tabular}{ l >{\ttfamily}l >{\ttfamily}l >{\ttfamily}l} \toprule
\textbf{Circuit Code} & $E_{J_1}$ [GHz] & $E_{J_2}$ [GHz]& $C$ [fF]\\
\midrule
\code{"JJ 1"} & 1.15 & 1.04 & 101\\ 
\code{"JJ 2"} & 6.79 & 1.00 & 38.1\\ 
Sycamore \cite{sycamore}& 18.0 & 18.0 & 108 \\ \bottomrule
\end{tabular}
\vspace{0.5em} 

\caption{\textbf{\code{"JL"} element values.} Optimized circuit element values for \code{"JL"} qubits compared to the heavy fluxonium qubit. The table presents the Josephson junction energy ($E_J$), inductance ($L$), and capacitance ($C$) for each circuit.}
\label{tab:JL_values}
\begin{tabular}{ l >{\ttfamily}l >{\ttfamily}l >{\ttfamily}l} \toprule
\textbf{Circuit Code} & $E_J$ [GHz] & $L$ [uH] & $C$ [fF]\\ \midrule
\code{"JL 1"} & 6.86 & 1.61 & 9.76\\ 
\code{"JL 2"} & 4.26 & 1.83 & 8.08\\ 
Heavy fluxonium \cite{heavy_fluxonium}& 3.39 & 1.23 & 40.4\\ \bottomrule
\end{tabular}
\end{table}

We utilized the Sherlock High-Performance Computing cluster at Stanford University to execute the optimization pipeline illustrated in Figure~\ref{fig:optimization_pipeline}. This pipeline optimized the loss function defined in Equation~\eqref{eq:qubit_search_2} using $100$ steps of the BFGS algorithm for each circuit instance. The hyperparameters related to the loss function weights were set to $\beta_\varphi = \beta_n = \beta_e = \beta_f= 1$, with maximum tolerances for external flux, charge, and element sensitivity set to $\mathcal{S}^*_\varphi = 0.1$, $\mathcal{S}^*_n = 0.02$, and $\mathcal{S}^*_e = 0.1$ respectively.

To determine the total truncation number $K$ (proxy for Hilbert space size; see Appendix~\ref{app:automatic_truncation}) for each circuit topology, we first initialized one hundred random circuits for each topology. We then plotted the flux spectrum for the first ten eigenfrequencies of these circuits vs total truncation number and selected the minimum truncation number required for the convergence of all one hundred circuits. This approach helps validate the output and methodology of the convergence test that we describe in Appendix~\ref{app:convergence_test}, and provides a total Hilbert space size that is sufficiently large to explore the search space, but not so large as to make optimization excessively slow. The maximum $K$ used in this experiment was $1.6 \times 10^4$, which was crucial for exploring larger circuits, such as those with four nodes.

The optimization results for some of the discovered qubits are summarized in Table~\ref{tab:search_results}. The table includes all necessary metrics for comparing optimized qubits, such as the total loss function $\mathcal{L}$, the number of gates $\mathcal{N}$, decoherence time $T$, and other relevant metrics. The optimized qubits are represented by their circuit code in a red string format. Additionally, we have included the corresponding circuits from the literature for each circuit topology next to the corresponding circuit code. This facilitates a more precise comparison between the discovered qubits and those previously built and documented in the literature. 

Since the estimated number of gates $\mathcal{N}$ was our primary optimization metric, and $\mathcal{N}$ depends on the decoherence time $T$ and the upper bound gate speed $\mathcal{G}$, we have plotted the qubits from Table~\ref{tab:search_results} in Figure~\ref{fig:search_results} based on these metrics. In the plot, the $x$-axis represents $\mathcal{G}$, and the $y$-axis represents $T$, both on a logarithmic scale. The dashed lines indicate constant contours of the gate number, expressed as $\mathcal{G}T=\text{const}$. Thus, qubits located towards the top-right corner of the plot correspond to those with a larger number of gates.

It is important to note that Figure~\ref{fig:search_results} does not display the metrics for flux, charge, and element sensitivities. For example, although the heavy fluxonium qubit's properties implies a high number of gates, it suffers from flux sensitivity, which our search has penalized. This issue is not present in our optimized \code{"JL"} circuits. Generally, all optimized circuits are designed to avoid such sensitivities, as specified in our loss function design in Equation~\eqref{eq:qubit_search_2}. 

To ensure that our optimization pipeline discovers circuits as intended, we focus on \code{"JJ"} circuits, which are analytically calculable and thus easier to understand. These circuits consist of two Josephson junctions, $E_{J_1}$ and $E_{J_2}$, which form an inductive loop. The junctions are not necessarily equal and are shunted by a single capacitor, $C$. Our search has identified two types of \code{"JJ"} qubits. The external flux spectrum and flux working points of these qubits, along with their literature counterpart, the Sycamore transmon \cite{sycamore}, are depicted in Figure~\ref{fig:optimal_circuits}a. The corresponding circuit element values are provided in Table~\ref{tab:JJ_values}.

The first discovered qubit, \code{"JJ 1"}, has $\varphi_\text{ext}^*$ close to zero and two junctions with values close to each other, combined with a large shunt capacitance to remove charge sensitivity. This suggests that our optimization pipeline has rediscovered the equivalent of the transmon qubit.

The second discovered qubit, \code{"JJ 2"}, has $\varphi_\text{ext}^* = \pi$, which is a less common working point for the transmon since it is very sensitive to external flux. However, our discovery pipeline circumvents this issue by identifying asymmetric Josephson junctions that reduce the flux sensitivity at $\varphi_\text{ext}^* = \pi$ \cite{engineering_guide}. Interestingly, both optimized qubits have an order of magnitude larger number of gates compared to their literature counterparts. It is important to clarify that our comparison of the optimized qubits and their literature counterparts, in terms of gate count, is intended solely to underscore the inherent potential of each circuit topology. For example, the Google Sycamore qubit may have been optimized for other functionalities, including superior two-qubit gates. Our intent is to emphasize the inherent capabilities of this type of qubit relative to other circuit topologies.

One of the key outcomes of our numerical experiments is the optimization of the \code{"JL"} circuits, which is a class of circuits that includes the fluxonium qubit. This topology consists of a Josephson junction $E_J$ and a superinductor $L$, which together form an inductive loop, and a capacitor $C$. The optimized \code{"JL"} circuits exhibit a number of gates that are an order of magnitude superior to Pop \textit{et al.}'s fluxonium \cite{pop_fluxonium} and heavy fluxonium \cite{heavy_fluxonium}. Not only do we predict a higher gate number metric, we also predict that the optimized qubits are significantly less sensitive to external fluxes. This reduced sensitivity mitigates the primary drawback of the heavy fluxonium circuit, which is prone to dephasing due to external flux interference.

To underscore the differences in flux sensitivity, we have plotted the flux spectrum of our optimized \code{"JL"} circuits alongside that of the heavy fluxonium in Figure~\ref{fig:optimal_circuits}b. The figure clearly illustrates that our optimized qubits are substantially less sensitive to external flux perturbations at their operational flux points. This innovative qubit design comprises a range of circuit element values that have not been investigated previously. For a detailed comparison, Table~\ref{tab:JL_values} provides a summary of the circuit element values for the optimized \code{"JL"} circuits, heavy fluxonium, and Pop's fluxonium.

Another interesting discovered qubit is the \code{"JJJ"} qubit depicted in Figure~\ref{fig:circuits}a, which we examine here to elaborate more on our optimization pipeline. This circuit consists of three Josephson junctions in a loop: $E_{J_1}$, $E_{J_2}$, and $E_{J_3}$, with capacitors $C_1$, $C_2$, and $C_3$. This circuit features two charge islands and one inductive loop, for which we must ensure that the islands have minimal charge dispersion and the loop is not flux-sensitive. The BFGS optimization of $100$ instances of this circuit topology is depicted in Figure~\ref{fig:optimal_circuits}c. The dashed black lines represent suboptimal circuits, while the solid red line represents the optimal (best-performing) circuit. Notably, the final results are sensitive to the initial choice of sampled circuit parameters. A significant majority of the initial sampled parameter choices converged to final loss values that were considerably higher than those of the optimal circuit, suggesting that the landscape is dominated by local optima, and depends strongly on the choice of initial condition.

The flux spectrum of the optimal circuit is shown in Figure~\ref{fig:optimal_circuits}d, and the charge spectrum for the excited states of the qubit is depicted in Figure~\ref{fig:optimal_circuits}e. The charge spectrum of the qubit shows that this qubit's first excited state is insensitive to the charge degree of freedom, as intended.

Another notable aspect of this qubit is that it has an order of magnitude higher number of gates compared to other circuits in the literature with the same topology, such as the flux qubit \cite{flux_qubit} and quantronium \cite{quantronium}. The eigenstates for the ground state and first excited state of the qubit, as a function of phase space coordinates, are plotted in the Hamiltonian potential landscape in Figure~\ref{fig:optimal_circuits}f. This suggests that this qubit has two-dimensional transmon-like eigenvectors localized in the minima of the circuit potential.

There are several qubit codes presented in Table~\ref{tab:search_results} and Figure~\ref{fig:search_results}, such as \code{"JLL"}, \code{"JL(JC)"}, and \code{"JJ(JC)"}, which propose intriguing qubits (see Figures~\ref{fig:circuits}c--e) that can be strong candidates as they support large $\mathcal N$ when optimized. However, we find that the optimized designs are ultimately nearly equivalent to previously discovered qubits with fewer elements. For example, the \code{"JLL"} qubit is composed of two inductors and one junction, forming an inductive loop. This qubit suggests a large number of gates comparable to \code{"JL"} qubit. However, the optimization achieves this number by shorting one of the inductors, reducing the circuit to fluxonium.

We observe a similar phenomenon with \code{"JL(JC)"}. This qubit also indicates a high value for $\mathcal{N}$. By examining the architecture of this qubit in Figure~\ref{fig:circuit_properties}a, we see that the circuit consists of a fluxonium capacitively coupled to a Cooper pair box. However, the optimizer achieves favorable numbers for this qubit by making the fluxonium off-resonant with the Cooper pair box, resulting in the lower levels of the eigensystem being dominated by the properties of fluxonium. 

The optimized \code{"JJ(JC)"} qubit suffers from the same issue. This circuit is essentially a flux-tunable transmon capacitively coupled to a non-flux-tunable Cooper pair box. The optimizer moves the Cooper pair box resonance away from the tunable transmon's resonance, leading to the lowest energy eigenstates of the total system being inherited from the Cooper pair box. Since this part of the system is not flux-tunable, we obtain a flux-insensitive spectrum for the qubit levels, and a qubit which is practically undesirable.

Further qubit designs such as \code{"JJL"}, \code{"JJJJ"}, \code{"JJJL"}, and \code{"JLJL"}, did not demonstrate significantly higher performance (by the considered metric) when compared to the optimized \code{"JL"} qubit. However, these qubits seemed to have spectra and physics that differed from fluxoniums and transmons. Moreover, given their greater number of nodes and richer physics, these qubits may offer additional degrees of freedom for encoding states, performing readout and realizing two-qubit interactions. These factors were not captured by the loss function employed in this study, presenting a potential area for future research.  Their circuit element values and their topologies are summarized in Figure~\ref{fig:circuits}. 

Overall, we demonstrate that our optimization pipeline can be utilized for search problems, specifically that of qubit discovery. This approach can lead to the identification of new qubits and improvements over the current state of qubits in the literature. Our pipeline adapts easily to other problems and can be improved by simply changing the loss function to suit the new problem. In summary, our search suggests the fluxonium qubit has the best overall  properties for the objective function we have considered.

\section{Outlook}
In this work, we have developed an optimization pipeline for optimizing properties of superconducting circuits. We have used this to search for superconducting qubit designs with the best performance. Our search was constrained to a specific range of element values and circuit topologies. We also assumed that the logical state of the qubit is encoded in the first excited state and ground state, leaving out for now alternative encodings such as bosonic~\cite{cat_code_paper} or dual-rail qubits~\cite{teoh2023dual, harry_levine_AWS}.

A second limitation in our approach is that, in estimating the upper bound of gate speed, we focused solely on the frequency spacing between states, without accounting for the coupling strength between them. However, coupling strength can also play a critical role in a practical setting where drive power is limited.

Additionally, our bounded estimate of gate speed does not account for certain considerations such as control constraints or sources of leakage induced by the drive source. Incorporating these quantities via a more robust modeling of the readout and control schemes used to interact with the qubit could increase numerical precision and help adapt the qubit design to more realistic scenarios.


Finally, we have only considered circuits with up to 4 nodes. Searching through circuits with a higher number of nodes required computational resource beyond those available to us currently. This is due to the exponential growth in memory requirements. To address this issue, more efficient representations of the circuit's Hilbert space, such as tensor networks~\cite{tensor_network} could be developed.

Our optimization pipeline can be directly applied to discover or optimize classes of superconducting circuits beyond qubits. For instance, this pipeline could be used to design coupler circuits that control the interaction between systems, such as toggling the interaction between two qubits on and off, enhancing qubit readout properties, or improving specific gate implementations for certain qubits. Additionally, it could be used to discover circuits that enhance the performance of quantum sensors or transducers.



\section{Online Presence}\label{sec:online_presence}
\code{SQcircuit} is a robust, BSD-3 licensed, open-source Python package that has been substantially enhanced in this iteration to include automatic differentiation capabilities. These capabilities are detailed in Section~\ref{sec:gradient} and are now integrated into the \code{SQcircuit} software as a new computational engine. The updated version is available on our GitHub repository, which also hosts a range of examples demonstrating the new features. Comprehensive documentation on how to install, update, and deploy these new features can be found on the official \code{SQcircuit} website. For developers and researchers interested in the optimization processes and qubit discovery algorithms discussed in Section~\ref{sec:qubit_discovery}, detailed source code and tutorials are available in a dedicated GitHub repository. We actively encourage the community to contribute to the ongoing development of \code{SQcircuit} through bug reports, comments, and improvements, or by initiating pull requests on our GitHub repositories. For a list of links to these resources, see Table~\ref{tab:links}.
\begin{table}[h!]
\caption{\textbf{Online Resources.} Links to the project's main website, source code repositories on GitHub, and example notebooks.}
\footnotesize
{
\def\arraystretch{1}
\begin{tabular}{@{}l@{}}
\toprule
\parbox{\columnwidth}{\code{SQcircuit} website and documentation:\\ \url{sqcircuit.org}}\\
\midrule
\parbox{\columnwidth}{GitHub repository for \code{SQcircuit} source code:\\ \url{github.com/stanfordLINQS/SQcircuit}}\\
\midrule
\parbox{\columnwidth}{GitHub repository for \code{SQcircuit} example notebooks:\\ \url{github.com/stanfordLINQS/SQcircuit-examples}}\\
\midrule
\parbox{\columnwidth}{GitHub repository for qubit discovery source code and examples:\\ \url{github.com/stanfordLINQS/Qubit-Discovery}}\\
\bottomrule
\end{tabular}}
\label{tab:links}
\end{table}

\section{Conclusion}\label{sec:conclusion}
In this paper, we introduced a generalized framework for gradient-based optimization of superconducting quantum circuits using the \code{SQcircuit} software package. This framework incorporates automatic differentiation, enabling the efficient computation of gradients for various circuit properties with respect to arbitrary circuit elements. We demonstrated this approach by applying it to the qubit discovery problem, allowing for the identification of qubit designs with improved performance metrics. The methodology presented in this work is not limited to qubit discovery but can be extended to other optimization problems within the realm of superconducting quantum hardware. By leveraging the capabilities of \code{SQcircuit}, researchers can explore a wide range of circuit configurations and optimization criteria, ultimately accelerating the development of advanced quantum technologies. We encourage the community to utilize and contribute to the open-source \code{SQcircuit} project, as detailed in Section~\ref{sec:online_presence}, to further enhance the tool's functionality and broaden its application scope.

\section*{Author Contributions}
T.R. and A.B. developed the initial theory for the backpropagation pass and wrote the initial implementation in \code{PyTorch}. T.R. and S.B. calculated the Hamiltonian derivatives. T.R., A.B., and S.B. implemented the optimization infrastructure in \code{SQcircuit}. T.R. and A.B. performed experiments on diagonalization convergence. T.R. implemented the BFGS algorithm. A.B., S.B., and T.R. performed experiments on Stanford's Sherlock cluster and analyzed the results. T.R. wrote the bulk of the manuscript, and used AI for grammatical editing of portions of the manuscript. A.B. and S.B. also wrote parts of the manuscript, including appendices. T.R. created \code{SQcircuit}’s website and established its online presence pipeline. A.H.S.-N. supervised the project from its inception. D.I.S. reviewed the manuscript and provided additional feedback. No AI tools were used at any stage in analytical derivations or in developing the code, unless explicitly stated. All authors reviewed the final manuscript and take full responsibility for its content.

\section*{Acknowledgments}
 We gratefully acknowledge funding from Amazon Web Services Inc., and NTT Research for their valuable financial and technical support throughout this project. We also recognize the support provided by the U.S. Government through the Air Force Office of Scientific Research and the Office of Naval Research under award number FA9550-23-1-0338, and Army Research Office (ARO)/Laboratory for Physical Sciences (LPS) Modular Quantum Gates (ModQ) program (Grant No. W911NF-23-1-0254), and the National Science Foundation CAREER award No.~ECCS-1941826. Some of the computing for this project was performed on the Sherlock cluster. We would like to thank Stanford University and the Stanford Research Computing Center for providing computational resources and support that contributed to these research results.

\newpage
\appendix

\section{PyTorch Computational Graph}\label{app:PyTorchExample}
\begin{figure}[t]
    \centering
    \includegraphics[width=0.5\textwidth]{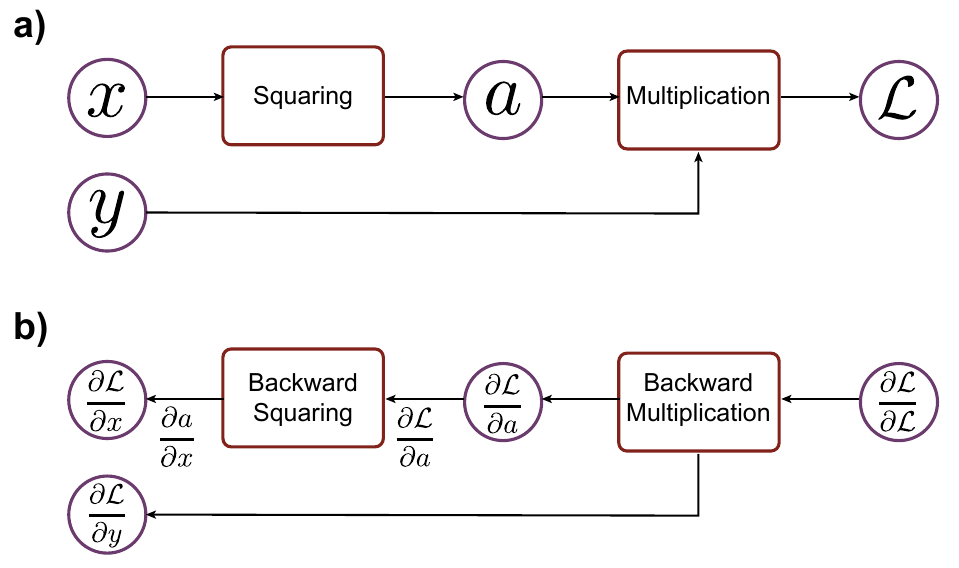}
    \caption{\textbf{Example of a computational graph. a)} The graph illustrates the sequence of operations and data flow during the forward pass. The input tensor $x$ is squared to produce the intermediate variable $a$, which is then multiplied by another input tensor $y$ to obtain the output loss $\mathcal{L} = x^2 y$. \textbf{b)} The backward pass through the computational graph, showing the gradients computation. The gradient of the loss with respect to each variable is computed by applying the chain rule. The gradients flow backwards through the graph: from the loss $\mathcal{L}$ to $y$ and $a$, and finally to $x$. The gradients of $x$ and $y$ are $\frac{\partial \mathcal{L}}{\partial x} = 2xy$ and $\frac{\partial \mathcal{L}}{\partial y} = x^2$, respectively.
        }
    \label{fig:example_graph}
\end{figure}

In this section, we explain the computational graph, forward pass, and backward pass functionalities of \code{PyTorch} with a simple example depicted in Figure~\ref{fig:example_graph}. \code{PyTorch}, an open-source machine learning library developed by Facebook's AI Research lab, provides a flexible and efficient platform for building deep learning models, performing tensor computations with GPU acceleration, and utilizing automatic differentiation to compute gradients. Its dynamic computational graph and ease of use make it widely popular in both academia and industry.

A computational graph is a directed graph where nodes represent operations or variables, and edges represent the flow of data. In \code{PyTorch}, computational graphs are created dynamically during the forward pass of a network, allowing for flexibility and ease of debugging.

In Figure~\ref{fig:example_graph}a, the computational graph consists of the following elements:
\begin{itemize}
    \item $x$: An input tensor with an initial value of 1.0, which requires gradient computation.
    \item Squaring: An operation that squares the value of $x$, producing an intermediate variable $a$.
    \item $a$: The result of squaring $x$.
    \item $y$: Another input tensor with an initial value of 1.0, which also requires gradient computation.
    \item Multiplication: An operation that multiplies $a$ by $y$, resulting in the output $\mathcal{L}$.
    \item $\mathcal{L}$: The final output, representing the loss function $\mathcal{L}=x^2y$.
\end{itemize}

The forward pass is the process of computing the output of a computational graph given an input. Each operation in the graph is applied sequentially to propagate the input through the network to produce the output. In the given example, the forward pass involves the following steps:
\begin{enumerate}
    \item $x$: Input tensor with $x=1.0$,
    \item $y$: Input tensor with $y=1.0$,
    \item $a$: Intermediate variable $a=x^2$,
    \item $\mathcal{L}$: Output loss $ \mathcal{L} = a \cdot y = x^2 \cdot y$.
\end{enumerate}

The corresponding \code{PyTorch} code for the forward pass is:
\begin{lstlisting}[language=Python]
import torch

x = torch.tensor([1.0], requires_grad=True)
y = torch.tensor([1.0], requires_grad=True)

a = torch.pow(x, 2)
L = torch.mul(y, a)
\end{lstlisting}

The backward pass is the process of computing the gradients of the loss function with respect to the input variables by applying the chain rule of calculus. This process, known as backpropagation, is essential for optimizing the parameters. In the given example that is also depicted in Figure~\ref{fig:example_graph}b, the backward pass involves the following steps:
\begin{enumerate}
    \item Compute the gradient of the loss with respect to the output $\mathcal{L}$:
    \[
    \frac{\partial \mathcal{L}}{\partial \mathcal{L}} = 1.
    \]
    \item Compute the gradient of the loss with respect to $y$:
    \[
    \frac{\partial \mathcal{L}}{\partial y} = x^2
    \]
    \item Compute the gradient of the loss with respect to $a$:
    \[
    \frac{\partial \mathcal{L}}{\partial a} = y.
    \]
    \item Compute the gradient of the intermediate variable $a$ with respect to $x$:
    \[
    \frac{\partial a}{\partial x} = 2x.
    \]
    \item Apply the chain rule to compute the gradient of the loss with respect to $x$:
    \[
    \frac{\partial \mathcal{L}}{\partial x} = \frac{\partial \mathcal{L}}{\partial a} \cdot \frac{\partial a}{\partial x} = y \cdot 2x.
    \]
\end{enumerate}

Using the \code{PyTorch} code provided, the backward pass is automatically handled by calling the \texttt{backward()} method on the loss tensor:
\begin{lstlisting}[language=Python]
L.backward()
\end{lstlisting}

After executing the backward pass, the gradients of $x$ and $y$ can be accessed via their \texttt{.grad} attributes:
\begin{lstlisting}[language=Python]
print(x.grad)  # Output: tensor(2.0)
print(y.grad)  # Output: tensor(1.0)
\end{lstlisting}
The output confirms that \texttt{x.grad} returns \texttt{2.0} and \texttt{y.grad} returns \texttt{1.0}, as expected.

\section{\texorpdfstring{\code{SQcircuit}}{SQcircuit} Tutorial}\label{app:Tutorial}
In this appendix, we will provide a short tutorial on using \code{SQcircuit} to calculate the gradient of circuit properties.

First, we import the \code{SQcircuit} and \code{PyTorch} library and use the \code{set_engine} function to switch the \code{SQcircuit} engine from \code{"NumPy"} to \code{"PyTorch"}:
\begin{lstlisting}[language=Python]
# Import the libraries
import SQcircuit as sq
import torch

# Switch to the PyTorch engine
sq.set_engine("PyTorch")
\end{lstlisting}
Next, we construct the circuit using \code{SQcircuit}. This is similar to the procedure described in \cite{SQcircuit}, but here we specify the elements or properties for which we want to calculate gradients by setting \code{requires_grad=True}. In the following example, we construct a transmon circuit and calculate the gradient of some of its properties with respect to its capacitor. We define the circuit as follows:
\begin{lstlisting}[language=Python]    
# Define the transmon elements
C = sq.Capacitor(
    12, "fF", requires_grad=True
)
JJ = sq.Junction(10, "GHz")
elements = {(0, 1): [C, JJ]}

# Define the transmon circuit
transmon = sq.Circuit(elements)
\end{lstlisting}
Here, we set \code{requires_grad=True} only for the capacitor and not for the Josephson junction, as we are interested in the gradient with respect to the capacitor, not the junction. Next, we set the truncation number for the only charge mode of the circuit and diagonalize the circuit to calculate its eigenfrequencies and eigenvectors:
\begin{lstlisting}[language=Python]
# Set truncation number
transmon.set_trunc_nums([100])

# Diagonalize the circuit
efreqs, evecs = transmon.diag(n_eig=10)
\end{lstlisting}
In contrast to the \code{"NumPy"} engine of \code{SQcircuit}, the variables \code{efreqs} and \code{evecs} are represented as \code{PyTorch} tensors instead of \code{NumPy} or \code{QuTiP} objects. Additionally, properties such as decoherence times, coupling operators, and element values are all represented by \code{PyTorch} tensors. This is essential for gradient calculation and maintaining an appropriate computational graph. It is also important to note that the precision of the eigenvector gradient is proportional to \code{n_eig}, denoted by $n_E$ in Equation~\eqref{eq:eig_vec_grad}. 

After diagonalizing the circuit Hamiltonian---an essential step before calculating any circuit properties---we aim to compute the gradient of the matrix element between the first excited state and the ground state with respect to the capacitor. The matrix element is defined as:
\begin{equation*}
    g_{10} = \frac{1}{c} \left| \bra{f_1} \hat{Q} \ket{f_0} \right|.
\end{equation*}
Here, $g_{10}$ is the matrix element, $c$ is the capacitance, $\ket{f_0}$ and $\ket{f_1}$ are the ground and first excited states respectively, and $\hat{Q}$ is the charge operator.
As $g_{10}$ takes the form of a scalar metric described in Equation~\eqref{eq:lossfunction}, we can use \code{SQcircuit} to calculate its gradient with respect to the capacitor:
\begin{lstlisting}[language=Python]
# Calculate the matrix element
g_10 = 1 / C.get_value() * torch.abs(
    evecs[1].conj()
    @ transmon.charge_op(1, "original")
    @ evecs[0]
)

# Backpropagation step
g_10.backward()

# Access the gradient
print(C.grad)
\end{lstlisting}
In the above code, \code{.backward()} propagates the gradient in the reverse direction (applying the chain rule) to compute the gradient of any property, here \code{g_10}, with respect to any element or property that has \code{requires_grad=True}. In this case, \code{C} is the only variable with gradient computation enabled. The units of the gradient in \code{SQcircuit} are generally given by:
\begin{equation*}
    \text{Unit}[\text{gradient}] = \frac{\text{Unit}[\text{property}]}{\text{Unit}[\text{element}]},
\end{equation*}
where, since the unit of the capacitor is farads, the unit of the gradient becomes the unit of the matrix element per farad. If the element is an inductor or junction, the default units for the gradient calculations are henry or hertz respectively (external flux and charge values are unitless). 

\section{Dephasing Time Gradient}\label{app:T2_gradient}
In \code{SQcircuit}, dephasing rates between eigenstates $i$ and $j$ are calculated for critical current, charge, and flux noise via the equation
\begin{equation*}
    \Gamma_{\varphi, \lambda} = \sqrt{|2\ln\omega_\text{low}t_\text{exp}|}\left\lvert A_\lambda\frac{\partial f_{ij}}{\partial \lambda}\right\rvert,
\end{equation*}
where $\omega_\text{low}$ is a fixed low-frequency cutoff, $t_\text{exp}$ is the measurement time, $A_\lambda$ is the noise amplitude, and $\lambda$ is the noisy parameter for the channel of interest. Specifically $\lambda = E_{J_j}$ for critical current noise, $n_{g_j}$ for charge noise, and $\varphi_{\text{ext}_j}$ for flux noise (summing over all instances present in the circuit).

Because this rate is dependent on the first derivative of $f_{ij}$, derivatives of $\Gamma_\varphi$ will depend on the second derivative of $f_{ij}$. In particular, for a parameter $x$ of the circuit,
\begin{align*}
    \frac{\partial \Gamma_{\varphi, \lambda}}{\partial x} &=\mathrm{sgn}\left(A_\lambda \frac{\partial f_{ij}}{\partial \lambda}\right)\sqrt{2\ln \omega_\text{low}t_\text{exp}} \nonumber \\
    &\quad \times \left(\frac{\partial A_\lambda}{\partial x}\frac{\partial f_{ij}}{\partial \lambda} + A_\lambda \frac{\partial^2 f_{ij}}{\partial x\partial \lambda}\right). \label{eq:grad_tphi}
\end{align*}

As discussed at the end of Section~\ref{sec:gradient}, because the first-order derivatives are implemented as a custom computational node, \code{PyTorch} cannot compute the second-order derivatives automatically. Therefore, we implement an additional computational node to calculate the gradient of $T_\varphi$. The first-order derivatives of $A_\lambda$ are all straightforward and are given in Table \ref{tab:a_grad}.

\newlength\q
\setlength\q{\dimexpr .33\columnwidth -\tabcolsep}
\begin{table}[t]
\caption{Derivatives of $A_\lambda$ for supported noise channels.
}\label{tab:a_grad}
\centering
\begin{tabular}{lll}    \toprule
Noise channel   & $A_\lambda$                       & $\partial A_\lambda/\partial x$\\ 
\midrule
Critical current & $A_\text{cc}E_J$ \footnotemark[1] & $A_\text{cc}\delta_{xE_J}$  \\
Charge           & $A_\text{ch}$ \footnotemark[2]   & 0\\
Flux noise       & $A_\text{flux}$ \footnotemark[3]  & 0\\ \bottomrule
\end{tabular}
\footnotetext[1]{The default value is $A_\text{cc} = 10^{-7}$.}
\footnotetext[2]{The default value is $A_\text{ch} = e \times 10^{-4}$.}
\footnotetext[3]{The default value is $A_\text{flux} = 2\pi \times 10^{-6}$.}
\end{table}

An expression for the second-order derivative of $f_i$ can be found by differentiating (\ref{eq:eig_val_grad}), which gives
\begin{equation}\label{eq:eig_val_secondgrad}
    \frac{\partial^2 f_i}{\partial x \partial \lambda} = \langle f_i \lvert \frac{\partial^2\hat{H}}{\partial x\partial \lambda}\rvert f_i\rangle + 2\Re \langle \frac{\partial f_i}{\partial x} \lvert \frac{\partial \hat{H}}{\partial \lambda}\rvert f_i \rangle.
\end{equation}

The partial derivatives for $\ket{f_i}$ are given by (\ref{eq:eig_vec_grad}), and the first-order derivatives of $\hat{H}$ in Table~\ref{tab:hamiltonian_gradient}, so the only novel feature of (\ref{eq:eig_val_secondgrad}) is the second-order Hamiltonian derivative. Luckily, these vanish for many combinations of $(x, \lambda)$. In Table~\ref{tab:h_squared}, we present all nonvanishing second-order derivatives for $\lambda \in \{E_{J_j}, n_{g_j}, \varphi_{\text{ext}_j}\}$ and $x \in \{c_i, l_i, E_{J_i}, \varphi_{\text{ext}_i}, n_{g_i}\}$. As in Table~\ref{tab:hamiltonian_gradient}, we assume all external fluxes are assigned to junctions.

\begin{table}
\caption{Nonvanishing second-order derivatives of the Hamiltonian required for computing gradients of $T_\varphi$.
}\label{tab:h_squared}
\footnotesize
\centering
\begin{tabular}{ll} \toprule
 Hamiltonian Derivative &  \;Original Basis \\ \midrule
 $\partial^2 \hat{H}/\partial \varphi_{\text{ext}_i}\partial E_{J_j}$  & $b_{ji}\sin\left(\frac{2\pi}{\Phi_0}{\bm{w}}^T_j{{\hat{\bm{\Phi}}}}+\bm{b}_j^T\bm{\varphi}_{\text{ext}}\right)$ \\  \midrule
$\partial^2 \hat{H}/\partial c_i\partial n_{g_j}$ & $-{\bm{e}_j}^T \bm{C}^{-1}\frac{\partial\bm{C}}{\partial c_i}\bm{C}^{-1}\hat{\bm{Q}}$ \\ \midrule
$\partial^2 \hat{H}/\partial n_{g_i}\partial n_{g_j}$ & ${\bm{e}_j}^T {\bm{C}}^{-1}{\bm{e}_i}$ \\ \midrule
$\partial^2\hat{H}/\partial E_{J_i}\partial \varphi_{\text{ext}_j}$ & $b_{ij}\sin \left(\frac{2\pi}{\Phi_0}{\bm{w}}^T_i{{\hat{\bm{\Phi}}}}+\bm{b}_i^T\bm{\varphi}_{\text{ext}}\right)$ \\ \midrule
$\partial^2\hat{H}/\partial \varphi_{\text{ext}_i}\partial \varphi_{\text{ext}_j}$ & $\begin{array}{l} \sum_{k\in \mathcal{S}_J} \Big[E_{J_k} b_{ki}b_{kj}\\\times\cos \left(\frac{2\pi}{\Phi_0}{\bm{w}}^T_k{{\hat{\bm{\Phi}}}}+\bm{b}_k^T\bm{\varphi}_{\text{ext}}\right)\Big] \end{array}$ \\ \bottomrule
\end{tabular}
\end{table}

\section{Upper Bound on Gate Speed}\label{app:gate_speed}
\begin{figure}[t]
    \centering
    \includegraphics[width=0.5\textwidth]{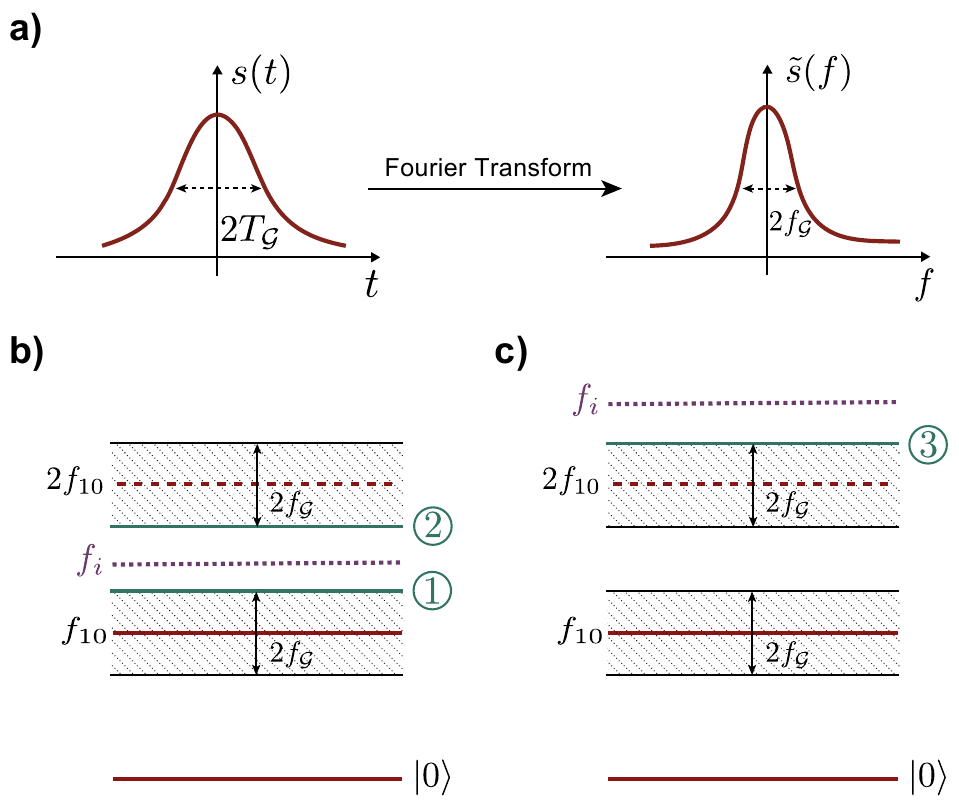}
    \caption{\textbf{Illustration of pulse signal and upper bound for Gate Speed.} \textbf{a)} The pulse signal $s(t)$ is depicted on the left with a width of $2T_\mathcal{G}$. The right figure shows the Fourier transform of the pulse $\tilde{s}(f)$, which has a width $2f_\mathcal{G} \propto 1/2T_\mathcal{G}$. \textbf{b)} The energy levels of the qubit with drive excitation frequencies $f_{10}$ and $2f_{10}$, along with a sideband box with a width of $2f_\mathcal{G}$. The eigenenergy $f_i$ of the $i$th excited state is lower than $2f_{10}$ and should fall between the green boundaries (1 and 2) to avoid excitation of the qubit states. \textbf{c)} The eigenenergy $f_i$ is higher than $2f_{10}$ and should be above the green boundary (3).}
    \label{fig:gate_speed}
\end{figure}
To implement a single-qubit gate, we need to drive the qubit with a time-dependent signal to achieve the desired unitary transformation that evolves the state of the qubit according to the requirements of the quantum gate. This involves externally coupling to the qubit and applying a signal with a frequency that matches the qubit frequency $f_{10}$, thereby enabling excitation between the two states of the qubit. The evolution of the system can be described by the total Hamiltonian, which is given abstractly as
\begin{equation*}
    \hat{H}(t) = \hat{H}_q + f(t)\hat{H}_d,
\end{equation*}
where $\hat{H}_q$ is the Hamiltonian of the qubit without any drive, $\hat{H}_d$ is the drive Hamiltonian that depends on the type of external coupling, and $f(t)$ is a time-dependent function typically expressed as
\begin{equation*}
    f(t) = s(t)\cos\left(2\pi f_{10} t\right),
\end{equation*}
where $s(t)$ is a pulse with a characteristic width of $2T_\mathcal{G}$, which defines the speed of the gate. In the Fourier domain of the pulse signal, as shown in Figure~\ref{fig:gate_speed}a, the pulse creates sidebands with a width of $2f_\mathcal{G} \propto 1/2T_\mathcal{G}$ around the excitation frequency $f_{10}$ and its harmonics $nf_{10}$, where $n$ is an integer. This is represented as a box around these excitation frequencies in Figure~\ref{fig:gate_speed}b. In this analysis, we avoid $n \ge 2$ because we assume the effect of the drive signal on these higher excitations is negligible. Therefore, increasing the speed of the gate (reducing the operation time) results in a wider box around the excitation frequencies. However, this increase is limited by the point at which higher excited states fall within these boxes, leading to excitations outside the computational space and a reduction in gate fidelity. To determine the permissible size of these boxes, we need to establish criteria to avoid overlap between the box edges and higher excited states. 

Assuming $f_i$ is the eigenenergy of the excited states for $i \ge 2$, we consider two cases based on the value of $f_i$ relative to $2f_{10}$. First, if $f_i < 2f_{10}$, we must ensure that $f_i$ falls within the green boundaries of 1 and 2 in Figure~\ref{fig:gate_speed}b. This requirement translates to the following conditions:
\begin{equation}\label{eq:gate_criteria_1}
    \begin{split}
        f_\mathcal{G} &< f_i - f_1,\\
        f_\mathcal{G} &< 2f_1 - f_i.
    \end{split}
\end{equation}
Second, if $f_i > 2f_{10}$, we must ensure that $f_i$ exceeds the green boundary of 3 in Figure~\ref{fig:gate_speed}c. This condition can be expressed as
\begin{equation}\label{eq:gate_criteria_2}
    f_\mathcal{G} < f_i - 2f_1.
\end{equation}
Therefore, we conclude that the upper bound for $f_\mathcal{G}$ should be the minimum of all criteria given in \eqref{eq:gate_criteria_1} and \eqref{eq:gate_criteria_2} for $i \ge 2$:
\begin{equation*}
    \mathcal{G} = \min_{i \ge 2} \left( f_i - f_1, \left| f_i - 2f_1 \right| \right).
\end{equation*}
\section{Diagonalization Convergence Test}\label{app:convergence_test}
When diagonalizing the Hamiltonian, the accuracy of the computed eigensystem is limited because we truncate the Hilbert space to a finite dimension. Moreover, during optimization we lack prior knowledge of the circuit's spectrum, so it is essential to establish a method to ensure the calculated results are trustworthy.

Assume the circuit has $n_N$ modes. To construct a finite-dimensional Hamiltonian, we restrict to a product Hilbert space formed from the $m_i$ lowest-energy eigenstates of each mode. (For harmonic modes, these are the Fock states $\ket{0}, \dots, \ket{m_i-1}$; for charge modes these are the charge number states $\ket{-(m_i - 1)/2}, \dots, \ket{0}, \dots, \ket{(m_i-1)/2}$.) We call each $m_i$ the dimension of mode $i$. For a harmonic mode this is equivalent to its truncation number; for a charge mode with truncation number $c_i$ this is given by $m_i = 2c_i-1$ (see Appendix~\ref{app:automatic_truncation}).

The total size of the Hilbert space is therefore
\begin{equation}
    D = \prod_{i=1}^{n_N} m_i.
\end{equation}
Thus, the Hamiltonian $\hat{H}$ we use is a $D \times D$ matrix, representing a truncated version of the true infinite-dimensional Hamiltonian. Ideally, the lowest-energy eigenenergies of the truncated Hamiltonian are close to those of the infinite-dimensional system. Traditionally, to test this, one increases each $m_i$ until the eigenvalues of interest stabilize. Once these eigenvalues no longer change, the diagonalization is considered converged. However, this iterative process is impractical during optimization. Therefore, an alternate method is required to verify the convergence of the eigenvalues and eigenvectors.


The eigenvalue problem for eigenenergy $f$ and eigenstate $\ket{f}$ is given by
\begin{equation*}
    \hat{H} \ket{f} = f \ket{f}.
\end{equation*}
Now, suppose the dimension of each mode is increased from $m_i$ to $m_i + t$, resulting in a Hilbert space of dimension $D'$. The eigenvalue problem for the corresponding $D' \times D'$ Hamiltonian $\hat{H}'$ is
\begin{equation*}
    \hat{H}' \ket{f'} = f' \ket{f'}.
\end{equation*}

If the Hamiltonian had already converged for the eigenvector so that $f = f'$, we expect that $\ket{f'}$ lies entirely in the original $D$-dimensional Hilbert space, and thus the components due to system expansion are zero. 

To check if this is the case, we embed the eigenvector $\ket{f}$ of the smaller system in the $D'$-dimensional system by concatenating it with zeros to match the size of $\ket{f'}$, resulting in $\ket{\tilde{f}}$. We then define the convergence error by
\begin{equation}\label{eq:convergence_error}
    \epsilon = 1 - | \braket{\tilde{f} | f} |^2.
\end{equation}
The criteria for convergence of the eigenvalue $f$ and eigenvector $\ket{f}$ is $\epsilon < \epsilon^*$. This has been implemented in \code{SQcircuit} and can be run on a diagonalized circuit as:
\begin{lstlisting}[language=Python]
epsilon_star = 1e-5
circuit.check_convergence(
    t=2,
    threshold=epsilon_star
)
\end{lstlisting}
Here, \code{circuit} is an object of the \code{Circuit} class in \code{SQcircuit}, and the \code{check_convergence()} function returns a boolean indicating the convergence state of the second excited state. The value of $\epsilon^*$ should be chosen small enough to ensure desired accuracy.

\section{Automatic Truncation Assignment}\label{app:automatic_truncation}
\begin{figure}
    \centering
    \includegraphics[width=1.0\linewidth]{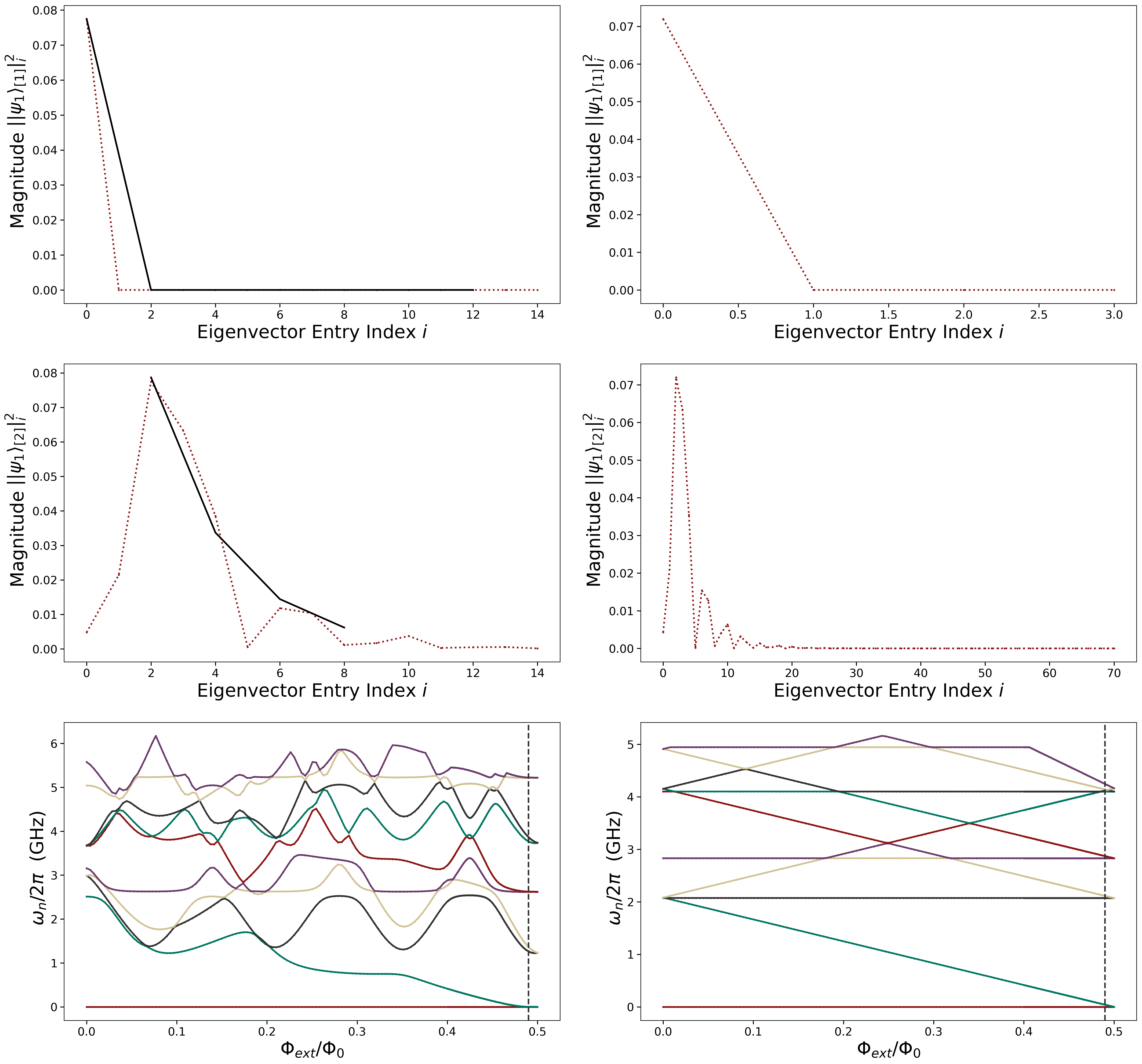}
    \caption{\textbf{Spectrum and eigenvector magnitudes for a \code{"JLJL"} circuit with truncation numbers allocated evenly (left column) or with the heuristic algorithm (right column).} In both cases, the product of truncation numbers is bounded by $K = 3000$. The top two rows show the mode magnitudes by index for each of the two harmonic modes present. The black lines in the first column are exponential fits to the eigenvector decay used in the heuristic algorithm. The bottom row shows the circuit spectrum for the lowest 10 states. On the left, stochasticity within the flux spectrum indicates that the assigned truncation numbers are insufficient to accurately compute the exact spectrum. On the right, the spectrum is converged below the same bound of $K$. The dashed black lines in the bottom row indicate the external flux used when applying the convergence test discussed in Appendix \ref{app:convergence_test}.}
    \label{fig:convergence_exponential_fit}
\end{figure}

\begin{figure}
    \centering
    \includegraphics[width=0.45\linewidth]{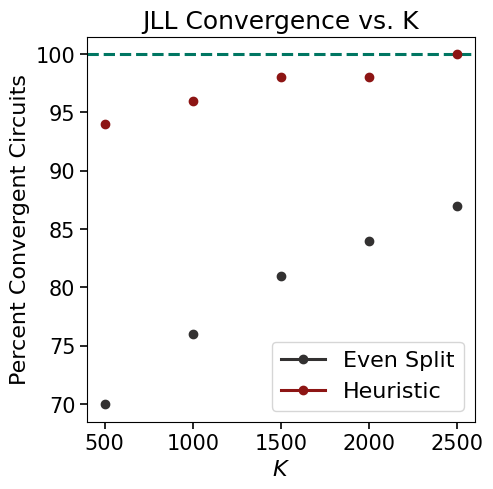}
    \includegraphics[width=0.45\linewidth]{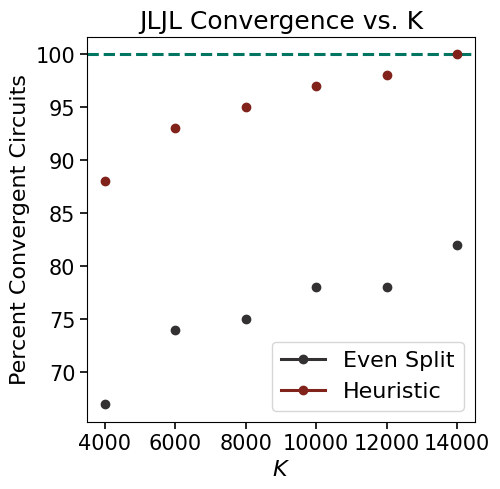}
    \caption{\textbf{Percentage of converged circuits at varying $K$ using even allocation and the heuristic algorithm.} 100 random circuits with the \code{"JLL"} (left) and \code{"JLJL"} topologies (right) were sampled, truncation numbers allocated for each fixed $K$, and convergence checked with the test (Appendix \ref{app:convergence_test}) and by eye.}
    \label{fig:convergence_comparison}
\end{figure}

\begin{figure}
\begin{algorithm}[H]
  \caption{Heuristic for truncation number assignment}\label{alg:truncation_heuristic}
  \begin{algorithmic}[1]
    \REQUIRE{Hamiltonian $\hat{H}$ expressed in a basis of $n_H$ harmonic and $n_C$ charge modes with $n_N\coloneqq n_H + n_C$. Bound $K$ on the product of truncation numbers and maximum charge truncation number $c_\text{max}$.} \\
    \STATE{Assign equal truncation numbers $\lfloor K^{1/n_N}\rfloor$ to each mode and diagonalize $\hat{H}$.}
    \FOR{each charge mode $j$}
    \IF{charge-spread estimation enabled}
    \STATE{Compute effective $E_{C_{j, \text{eff}}}$ and $E_{J_{j, \text{eff}}}$ as in Eq. \eqref{eq:charge_cob}.}
    \STATE{Calculate approximate width of ground state $\sigma_j =(E_{J_{j, \text{eff}}}/8E_{C_{j, \text{eff}}})^{1/4}$.}
    \STATE{Set $c_j=\min{(c_\text{max},\lfloor 3\sigma_j\rfloor)}$.}
    \ELSE
    \STATE{Assign fixed charge truncation $c_j = c_\text{max}$.}
    \ENDIF
    \ENDFOR
    \STATE{Compute remaining available Hilbert space dimension $K_h = K/\prod_{j=1}^{n_C}c_j$.}
    \FOR{each harmonic mode $i$}
    \STATE{For the highest eigenvector of interest $\ket{\psi}$, determine the highest index peak $d_i$ in the eigenvector's squared magnitudes $|\ket{\psi}_{[i]}|^2$.}
    \STATE{Fit an exponential decay curve of the form $f(x)\propto \exp{(-\alpha_i (x-d_i))}$ to the exponential decay in magnitudes following $d_i$.}
    \STATE{Calculate preliminary truncation numbers $\{h_i\}$ using Eq. \eqref{eq:preliminary_cutoffs}.}
    \STATE{Shift each truncation number $h_i\gets h_i+d_i$ and renormalize using Eq. \eqref{eq:renormalize_h}.}
    \ENDFOR
    \RETURN{Truncation numbers $\{c_j\}_{j=1}^{n_C}$ for charge modes and $\{h_i\}_{i=1}^{n_H}$ for harmonic modes.}
  \end{algorithmic}
\end{algorithm}
\end{figure}

In this appendix, we present a general algorithm to choose truncation numbers for charge and harmonic modes. The algorithm is designed to minimize the total dimension of the composite Hilbert space while maximizing the accuracy of computed results of interest. By combining this algorithm with the convergence test described in the previous appendix, we ensure that our search algorithm is both computationally efficient and maximizes the region of parameter space explored, given fixed computational resources.

As in the previous appendix, assume the circuit has $n_N$ total modes, which are divided into $n_H$ harmonic modes and $n_C$ charge modes. We fix an upper bound $K$ on the product of the modes' truncation numbers, which acts as a proxy for the total Hilbert space dimension and consequently for the total memory and computational resources required. For instance, the complexity of sparse diagonalization scales as $O(K^2)$. Following the implementation in \code{SQcircuit}, a harmonic mode with truncation number $h_i$ retains Fock states $\ket 0,\dots,\ket{h_i-1}$, while a charge mode with truncation number $c_j$ encompasses the $2c_j-1$ charge states $\ket{-(c_j-1)},\dots,\ket{c_j-1}$.

The algorithm discussed here outputs truncation numbers $h_i$ for each harmonic mode and $c_j$ for each charge mode satisfying
\begin{equation}
    \left(\prod_{i=1}^{n_H}h_i\right)\left(\prod_{j=1}^{n_C}c_j\right)\leq K
    \label{eq:K_upper_bound}.
\end{equation}

A na\"ive approach to this problem is to allocate truncation numbers evenly, by setting
\begin{equation}\label{eq:even_alloc}
    h_i = c_j\coloneqq\lfloor K^{1/n_N}\rfloor,\quad \text{for all $i, j$.}
\end{equation}
Since we diagonalize the circuit in the tensor-product basis of weakly coupled modes, the probability density of low-energy eigenstates is concentrated in low-energy excitations of the distinct modes. However, depending on the circuit, these eigenstates may have more excitations in some modes than others. Uniformly setting truncation numbers allocates excessive computational resources to modes in which the eigenstates have relatively few excitations, and deprives computational resources from those modes in which the eigenstates have many.

To more efficiently allocate truncation numbers, we aim to estimate how much weight the eigenstates have in each mode. For charge modes, we have an explicit approximation. The Hamiltonian in \texttt{SQcircuit} for these modes takes the form
\begin{equation}
    \hat{H}^\text{ch} = \frac{1}{2}\hat{\bm{Q}}^{\text{ch}, T}(C^\text{ch})^{-1}\hat{\bm{Q}}^\text{ch} + \sum_{k \in \mathcal{S}_J}E_{J_k}\cos(\bm{w}_k^T\hat{\bm{\varphi}} + \bm{b}_k^T\bm{\varphi}_\text{ext}).
\end{equation}
After a change of basis which diagonalizes $(C^\text{ch})^{-1}$, and expanding the cosine terms to second order, we have
\begin{equation}\label{eq:charge_cob}
    \hat{H}^\text{ch} = \sum_{j=1}^{n_C}\left[4E_{C_j, \text{eff}}\hat{\tilde{Q}}^2_j + E_{J_j, \text{eff}}\varphi_j^2\right] + \sum_{k \in \mathcal{S}_J}A_{ij}\varphi_i\varphi_j,
\end{equation}
where we have discarded gate charges and external fluxes. Further ignoring the coupling terms $A_{ij}\varphi_i\varphi_j$, this is equivalent to $n_C$ uncoupled harmonic oscillators, with ground state of approximate width $\sigma\coloneqq(E_{J, \text{eff}}/8E_{C, \text{eff}})^{1/4}$ in the number basis \cite{scqubit}. This estimate suggests choosing $c_j=\min(c_\text{max}, \lfloor\Delta\sigma_j\rfloor)$, where the number of standard deviations $\Delta$ retained determines the accuracy of calculated values. We implemented this option with $\Delta=3$ and $c_\text{max}=12$, and used this to produce the example in Fig.~\ref{fig:convergence_exponential_fit}. For the optimization runs reported in Section~\ref{sec:qubit_discovery}, we instead assigned a fixed charge mode truncation number $c_j=c_\text{max}=12$ (23 charge states) to circuits that also include harmonic modes. Circuits comprised only of charge modes used the even allocation of truncation numbers laid out in Eq.~\eqref{eq:even_alloc}.

After choosing truncation numbers for the charge modes, the dimension of the harmonic mode subspace is bounded by
\begin{equation}\label{eq:k_h_upper_bound} K_h := K\big/\prod_{j=1}^{n_C} c_j \end{equation}
For the harmonic modes, we do not have an \emph{a priori} estimate. However, because the Hamiltonian matrix has exponential off-diagonal decay, the eigenvectors generically have exponentially decaying weight in higher excitations for each mode \cite{benzi2016exploiting}. By comparing the decay \emph{rates} between different harmonic modes, we can infer the relative distribution of excitations between them.

We begin by uniformly allocating truncation numbers as in \eqref{eq:even_alloc} and diagonalizing the circuit Hamiltonian. Given a state written in the product basis of modes $1, \dots, n_N$ as
\[ \ket{\psi} = \sum_{l_1, \dots, l_{n_N}}c_{l_1 \cdots l_{n_N}}\ket{l_1}\cdots \ket{l_{n_N}}, \]
we define the state's component in mode $i$ by
\begin{equation}\label{eq:component-in-mode} |\ket{\psi}_{[i]}|^2 := \sum_{l_i} \max_{l_1 \cdots \widehat{l_i} \cdots l_{n_N}}\{|c_{l_1 \cdots l_i \cdots l_{n_N}}|^2\}\ket{l_i}. \end{equation}
This is a $\lfloor K^{1/n_N}\rfloor$-length column vector, where the $l_i$th entry indicates the maximum weight of $\ket{\psi}$ in the $l_i$th excitation of the $i$th mode. 

Using this definition, we extract the magnitudes $\{|\ket{\psi}_{[i]}|^2\}_{i=1}^{n_H}$ for some proxy eigenvector $\ket{\psi}$, conventionally the first excited eigenstate. Typically, $|\ket{\psi}_{[i]}|^2$ will be dominated by a small number of excitations, with an exponential decay away from them. We identify the largest index $d_i \in [1, \lfloor K^{1/n_N}\rfloor]$, corresponding to the highest-energy excitation in mode $i$, for which $|\ket{\psi}_{[i]}|^2$ has a peak. A peak is defined by $|\ket{\psi}_{[i]}|^2_l$ being lower for $l < d_i$ and $l>d_i$, as well as having a magnitude above a minimum threshold. Given that peak, we fit an exponential curve to the tail of the distribution, allowing us to extract the relative decay rates at the tail end of each mode. 

Provided such a fit $|\ket{\psi}_{[i]}|^2_l \sim A_i e^{-\alpha_i(l - d_i)}$, we aim to allocate more mode numbers to circuits with smaller $\alpha_i$ because a slower decay indicates there are more excitations in that mode. To quantify the tradeoff, we minimize the convergence error $\epsilon$ defined in Equation~\eqref{eq:convergence_error}. Assuming exponential decay, for large $\alpha$ then
\begin{equation}
\epsilon \sim \prod_{i=1}^{n_H} e^{-\alpha_i h_i}.
\end{equation}
Minimizing this subject to the constraint $\prod_{i=1}^{n_H} h_i \leq K_h$ sets the truncation numbers to
\begin{equation}\label{eq:preliminary_cutoffs}
    h_i = \left\lfloor \frac{1}{\alpha_i}\Big(K_h\prod_{j=1}^{n_H} \alpha_j\Big)^{1/n_H} \right\rfloor.
\end{equation}

While this accounts for the relative decay rates, we also need to account for the relative positions of each harmonic mode's peak $d_i$. To address this, we adjust each truncation number by the relative position of each mode's peak, taking $h_i\gets h_i+d_i$. To ensure that constraint on dimension remains satisfied, we then renormalize by setting 
\begin{equation}\label{eq:renormalize_h}
    h_i \gets h_i\left(\frac{K_h}{\prod_{i=1}^{n_H}h_i}\right)^{1/n_H}.
\end{equation}

This yields the final harmonic number cutoffs $\{h_i\}_{i=1}^{n_H}$, which are applied in conjunction with the charge cutoffs $\{c_j\}_{j=1}^{n_C}$. The overall process is summarized in Algorithm \ref{alg:truncation_heuristic}.

In practice, this heuristic algorithm represents a substantial improvement over the even allocation of truncation numbers defined by Eq. \eqref{eq:even_alloc}. One example is shown in Fig. \ref{fig:convergence_exponential_fit}, which demonstrates the improvement afforded by the heuristic allocation for a fixed $K=3000$. The left half of the plot shows an even truncation allocation of $h_1=h_2=14$, where the upper plot shows a faster decaying mode and the lower plot a relatively slower decaying mode. After applying the heuristic algorithm, a higher mode cutoff is allocated to the slower decaying harmonic mode leading to $h_1=4$ and $h_2=71$. For the same bound $K$, the flux spectra on the bottom row demonstrate a clear difference between the non-convergent eigenvalues on the left and the convergent values on the right.

To check the performance of this algorithm, we compared even allocation to the truncation numbers from Algorithm \ref{alg:truncation_heuristic} by randomly sampling a set of 100 circuits of multiple circuit types and testing the proportion of convergent circuits produced by both. These results for circuits of type \code{"JLL"} (two harmonic modes) and \code{"JLJL"} (two harmonic modes and one charge mode) are presented in Fig. \ref{fig:convergence_comparison}, which demonstrates that the heuristic algorithm produces a higher proportion of convergent circuits than the even distribution for a wide range of values for $K$. As this trend continues even to 100\% convergence for a wide class of circuits, we have strong evidence that the heuristic generally helps allocate computational resources more efficiently and explore a wider swath of parameter space with the same computational constraints. In these tests, charge modes were assigned the same even split of $K$ under both allocations, so this comparison isolates the effect of the heuristic algorithm used to assign harmonic modes.

\bibliographystyle{quantum}
\bibliography{main}

\end{document}